\documentclass[]{aa}  

\usepackage{graphicx}
\usepackage[symbol]{footmisc}
\usepackage{txfonts}
\usepackage[colorlinks=true, citecolor=blue]{hyperref}
\usepackage{ulem}
\usepackage{caption}
\newcommand{\micron}{$\mu$m}

\newcommand{\uv}{{\it uv}}
\newcommand{\code}[1]{{\color{blue} \texttt{#1}}}
\newcommand{\pack}[1]{{\color{magenta} \texttt{#1}}}

\renewcommand{\thefootnote}{\fnsymbol{footnote}}
\usepackage[dvipsnames]{xcolor}
\definecolor{orcidlogocol}{HTML}{0000FF}

\begin{document}

   \title{The dusty heart of Circinus }

   \subtitle{II. Scrutinizing the \textit{LM}-band dust morphology using MATISSE \thanks{This work makes use of ESO Programmes 099.B-0235, 0101.B-0446, 0101.C-092, 0104.B-0064(A), 0104.B-0127(A), 105.205M.001, and 106.214U.002.}}

    \author{
    \href{https://orcid.org/0000-0002-1272-6322}{\textcolor{orcidlogocol}{J. W. Isbell}}
    \inst{1}
        \and
    J.-U. Pott \inst{1}
        \and
    K. Meisenheimer$^{\dagger}$ \inst{1}
        \and
    M. Stalevski \inst{2,3}
        \and
    K. R. W. Tristram \inst{4}
        \and
    J. Leftley \inst{5}
        \and
    D. Asmus \inst{6,7}
        \and
    G. Weigelt \inst{8}
        \and
    V. G\'amez Rosas \inst{9}
        \and 
    R. Petrov \inst{5}
        \and
    W. Jaffe \inst{9}
        \and 
    K.-H. Hofmann \inst{8}
        \and
    T. Henning \inst{1}
        \and
    B. Lopez \inst{5}
    }

   \institute{
   Max-Planck-Institut f\"ur Astronomie (MPIA), K\"onigstuhl 17, 69117 Heidelberg, Germany\\ \email{isbell@mpia-hd.mpg.de}
   \and
   Astronomical Observatory, Volgina 7, 11060 Belgrade, Serbia
   \and 
   Sterrenkundig Observatorium, Universiteit Gent, Krijgslaan 281-S9, Gent, 9000, Belgium
  \and 
  European Southern Observatory, Alonso de C\'ordova 3107, Vitacura, Santiago, Chile
  \and 
  Laboratoire Lagrange, Universit\'e C\^ote d'Azur, Observatoire de la C\^ote d'Azur, CNRS, Boulevard de l'Observatoire, CS 34229, 06304 Nice Cedex 4, France
  \and 
  Gymnasium Schwarzenbek, 21493 Schwarzenbek, Germany
  \and 
  Department of Physics \& Astronomy, University of Southampton, Southampton, SO17 1BJ, UK
  \and
  Max-Planck-Institut f\"ur Radioastronomie, Auf dem H\"ugel 69, D-53121 Bonn, Germany
  \and 
  Leiden Observatory, Leiden University, Niels Bohrweg 2, NL-2333 CA Leiden, The Netherlands
}

   \date{Received 28 June 2023; accepted 1 September 2023}

 
  \abstract{
  In this paper we present the first-ever $L$- and $M$-band interferometric observations of Circinus, building upon a recent $N$-band analysis. We used these observations to reconstruct images and fit Gaussian models to the $L$ and $M$ bands. Our findings reveal a thin edge-on disk whose width is marginally resolved and is the spectral continuation of the disk imaged in the $N$ band to shorter wavelengths. Additionally, we find a point-like source in the $L$ and $M$ bands that, based on the $LMN$-band spectral energy distribution fit, corresponds to the $N$-band point source. We also demonstrate that there is no trace of direct sightlines to hot dust surfaces in the circumnuclear dust structure of Circinus. By assuming the dust is present, we find that obscuration of A$_{\rm V} \gtrsim 250$ mag is necessary to reproduce the measured fluxes. Hence, the imaged disk could play the role of the obscuring "torus" in the unified scheme of active galactic nuclei. Furthermore, we explored the parameter space of the disk + hyperbolic cone radiative transfer models and identify a simple modification at the base of the cone. Adding a cluster of clumps just above the disk and inside the base of the hyperbolic cone provides a much better match to the observed temperature distribution in the central aperture. This aligns well with the radiation-driven fountain models that have recently emerged. Only the unique combination of sensitivity and spatial resolution of the VLTI allows such models to be scrutinized and constrained  in detail. We plan to test the applicability of this detailed dust structure to other MATISSE-observed active galactic nuclei in the future.}

   \maketitle
%
\renewcommand{\thefootnote}{\fnsymbol{footnote}}
\section{Introduction}
\footnotetext[2]{Deceased February 4, 2023}
 Active galactic nuclei (AGNs) are thought to play a crucial role in the formation and evolution of their host galaxies. Moreover, understanding the dust in the vicinity of supermassive black holes is key to understanding how AGNs are fed and how they interact with their hosts. The dust traces dense molecular gas, which feeds the accretion disk (AD). Large, obscuring dusty structures are thought to be responsible for both funneling material toward the AD and for causing the apparent differences between Seyfert 1 and Seyfert 2 galaxies. In the original unified model of AGNs \citep{antonucci1993, urry1995,netzer2015}, a central obscuring torus of dust is oriented such that the broad-line region (BLR) of the AGN is directly visible (Seyfert 1) or such that its observation is blocked by the torus (Seyfert 2; hereafter Sy2). So in order to fully understand the accretion process and the life cycle of an AGN, one must understand the parsec-scale dust structures surrounding it.

\renewcommand{\thefootnote}{\arabic{footnote}}
The so-called torus comprises several key features that vary in temperature from $<100$ K to $1500$ K and scale from tenths of a parsec to tens of parsecs.
The inner edge is the radius at which radiation from the AD causes the dust to sublimate. The sublimation radius is dependent on both the luminosity of the AD and the dust composition, typically $\sim 0.1$ pc for a $L \sim 1 \times 10^{10} L_{\odot}$ AGN with dust evaporating at $1500$ K \citep[see, e.g.,][]{barvainis1987}. 
Beyond the sublimation zone, it is thought that a dense disk or torus of material is responsible for hiding the BLR in Sy2 AGNs, for feeding the AD, and for reflecting X-rays. Previous mid-infrared (MIR) interferometric studies revealed that many ``tori'' have an additional component in the form of a polar extension \citep[see, e.g.,][]{honig2012,burtscher2013,lopez-gonzaga2016,leftley2018}, the Circinus Galaxy chief among them \citep[][]{tristram2007c,tristram2014}. The polar component is thought to be a radiation-driven outflow \citep[e.g.,][]{wada2012,wada2016}, and it can represent a key mechanism of AGN feedback. This is called the fountain model, and it was shown that it can potentially explain the MIR polar extension \citep{schartmann2014}. \citet{stalevski2017,stalevski2019} used a model combining a dusty hollow cone and a thin disk to reproduce the spectral energy distribution (SED) and morphology of the torus in the Circinus Galaxy (hereafter Circinus). 
A key finding of SED fits to nearby AGNs as well as comparisons to radiative transfer (RT) models is that the dust in the central structures (and particularly in the wind) must be clumpy, allowing dust to reach high temperatures and exhibit ``blue'' spectra even at large distances from the AD \citep[][]{krolik1988,nenkova2008a, schartmann2008, honig2017,martinez-paredes2020,isbell2021}. 
The exact nature of these components and how they are connected to each other and to the host galaxy remains an open question.  A holistic model of the central dust distribution is shown in \citet{izumi2018}, but only the resolution offered by infrared interferometry can probe the subparsec details of the dust near the active nucleus.

The Multi AperTure mid-Infrared Spectro-Scopic Experiment (MATISSE) is the second-generation MIR interferometer on the Very Large Telescope Interferometer (VLTI) at the European Southern Observatory (ESO) Paranal site \citep{lopez2014,lopez2022}. MATISSE combines the light from four unit telescopes (UTs) or four auxiliary telescopes (ATs), measuring six baselines in the \textit{L}, \textit{M}, and \textit{N} bands simultaneously. MATISSE furthermore introduces closure phases to MIR interferometry. The combination of the phase measurements on any three baselines, $\phi_{ijk} \equiv \phi_{ij} + \phi_{jk} - \phi_{ik}$, is called the closure phase; this summation cancels out any atmospheric or baseline-dependent phase errors \citep[][]{jennison1958, monnier2003}. Closure phases are crucial for imaging because they probe the spatial distribution of target flux and because they are unaffected by atmospheric turbulence. Recent imaging studies of NGC\,1068 with VLTI/GRAVITY \citep{gravitycollaboration2020} and VLTI/MATISSE \citep{gamezrosas2022} have illustrated the power of this approach in revealing new morphological details and spatially resolved temperature measurements of the circumnuclear dust. 

Circinus is of particular interest as it is one of the closest Sy2 galaxies \citep[at a distance of 4.2 Mpc;][]{freeman1977, tully2009} and the second brightest in the MIR (only fainter than NGC 1068). 
Circinus is a prototypical Sy2 galaxy, exhibiting narrow emission lines \citep[][]{oliva1994,moorwood1996} and an obscured BLR \citep{oliva1998, ramosalmeida2016}, as well as bipolar radio lobes \citep[][]{elmouttie1998} and an optical ionization cone \citep[][]{marconi1994,maiolino2000,wilson2000,mingozzi2019,kakkad2023}. Additionally, Circinus exhibits a Compton-thick nucleus and a reflection component in X-rays \citep[][]{matt1996,soldi2005,yang2009,arevalo2014}. Finally, inflows, outflows, and spiral arms have been observed in CO down to $\sim5$ pc scales \citep[][]{curran1998, izumi2018,tristram2022}, further indicating the complexity of the central structures.

Circinus was recently imaged for the first time with MATISSE in the $N$ band \citep{isbell2022}.
These images revealed a dust disk roughly aligned with the water maser emission \citep{greenhill2003}, as well as warm ($\sim 250$~K) large-scale ($\gtrsim 100$ mas) emission roughly orthogonal to the disk, similar to previous results with the first-generation MIR interferometer, MIDI \citep{tristram2014}. 
The orientation of the large-scale emission's major axis was found to differ significantly from the optical ionization cone central angle (PA$_{\rm opt.}$ = $-45^{\circ}$ vs. PA$_{\rm dust} = -73^{\circ}$), and the MATISSE images revealed flux enhancements along the position angle (PA) of the optical ionization cone (PA$_{\rm opt.}$ = $-45^{\circ}$). Previous modeling work by \citet{stalevski2017,stalevski2019} has indicated that this enhanced dust emission may come from an edge-brightened outflow cone.   

The proximity and declination of Circinus (at around $-60^{\circ}$) make it an ideal target for imaging with MATISSE, as it provides high spatial resolution (10 mas $=$ 0.2 pc) and because its nearly circular $uv$ tracks aid in the production of high fidelity reconstructions. 
MATISSE provides the first MIR measurements of the closure phase, which sample the (a)symmetry of a source and are crucial for image reconstruction. Previous analyses relied on Gaussian model fitting, which is a smooth, simplified representation of the source emission; interferometric image reconstruction has the potential to build on these results through model-independent sampling of the source structure. In \citet[][hereafter Paper I]{isbell2022} we present the first image reconstructions of Circinus in the $N$ band. In this work we extend the analysis to the \textit{LM} bands and further consider RT models that represent the source. 

This paper is structured as follows. We present the observations used in this work in Sect. \ref{sec:obs} along with the necessary data reduction steps. In Sect. \ref{sec:imaging} we explain our modeling and imaging methods and show the results of each. In Sect. \ref{sec:bb} we consider the spatially resolved temperatures of the recovered images and models. Finally, we discuss the implications of this work in Sect. \ref{sec:discussion} and conclude in Sect. \ref{sec:conc}.

\section{Observations and data reduction}
\label{sec:obs}
\subsection{MATISSE observations}
\label{sec:circ_obs}
\begin{figure}
    \centering
    \includegraphics[width=0.5\textwidth]{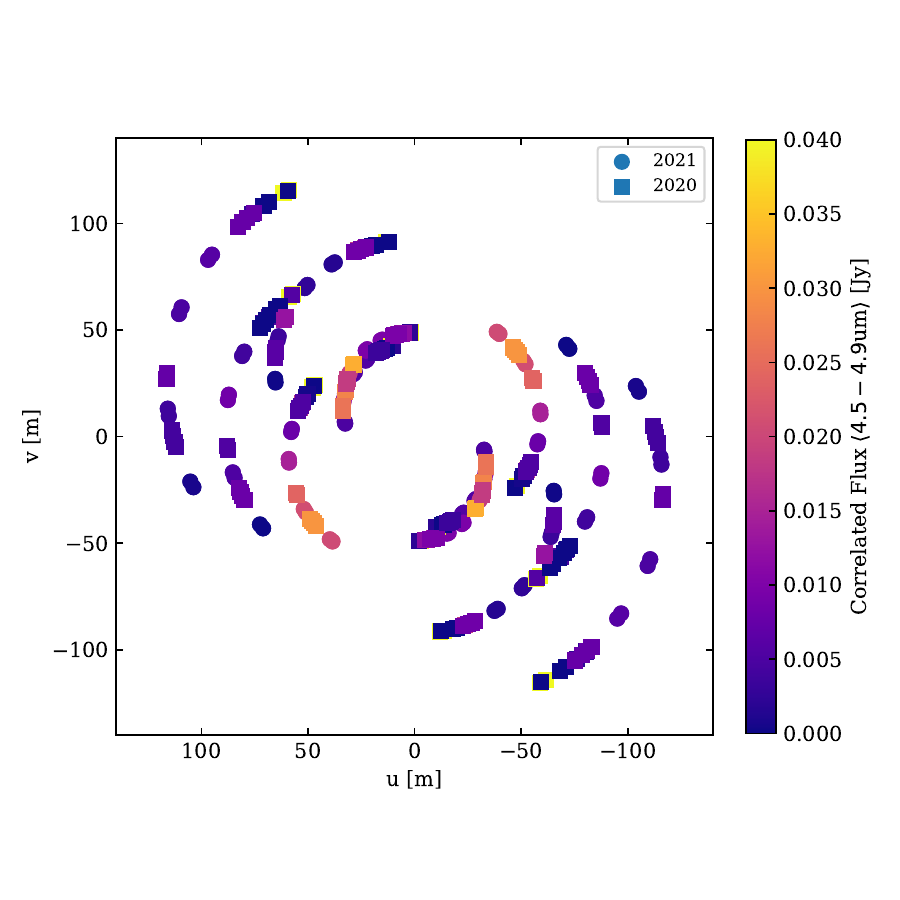}
    \caption{MATISSE \textit{uv} coverage from all 25 exposure cycles. Squares denote observations taken in 2020, and circles represent observations from 2021. The mean correlated flux between 4.5 and 4.9~$\mu$m is used as the color scale. North is up, and east is to the left.}
    \label{fig:uv}
\end{figure}

The MATISSE observations of Circinus were carried out on 13--14 March 2020, 27 February 2021, and 31 May 2021 as part of guaranteed time observations. 
Data were taken with low spectral resolution in both the $LM$- (3--5 $\mu$m) and $N$ bands (8--13 $\mu$m). The observations were taken using the UT configuration, with physical baselines ranging from 30 to 140 m. The observations are described in more detail in Paper I. 
We show the combined \uv\  coverage of all the observations in Fig. \ref{fig:uv}.

On each night, we observed the calibration star HD120404 ($F_{12\mu{\rm m}} = 13$ Jy) directly before and/or after the Circinus observations. 
This star serves as a spectral calibrator, as an instrumental phase calibrator, and as an instrumental visibility calibrator. It has a MIR spectrum given by \citet{vanboekel2004}, and its diameter is given as 2.958 mas in \citet{cruzalebes2019}. During the February and May 2021 observations, we observed secondary calibrators, HD120913 ($F_{12\mu{\rm m}}=5.7$ Jy) and HD119164 ($F_{12\mu{\rm m}}=1.2$ Jy) in order to perform cross-calibration and closure phases accuracy checks. All Circinus and calibrator observations entering this analysis are given in Table I of Paper I.

\subsection{MATISSE \textit{LM}-band data reduction and calibration}
\label{sec:circl_reduction}
The $LM$-band data for Circinus and the calibrators were reduced using both the official data reduction software (DRS) version 1.5.1 and custom scripts. We reduced the data both coherently and incoherently. For the coherent reduction we used the flags \code{corrFlux=TRUE} and \code{coherentAlgo=2} in order to produce correlated fluxes. For the incoherent reduction we used the flag \code{corrFlux=FALSE}.
In both cases we used spectral binning 11 px (= 0.5 \micron) and the default values for all other parameters. The DRS is not optimized for coherent reduction of the $LM$ bands, but coherent integration is necessary for faint sources \citep[see, e.g., the flux limits given in][]{lopez2022}. The above settings (both coherent and incoherent) resulted in strange spectra with (1) no $M$-band flux and (2) sharp emission features at 3.7 \micron; but analysis of the intermediate products (specifically, the cleaned interferogram) found neither of these features. It was found that a bias floor was present in the DRS-reduced data. This is likely due to the fringe search being optimized for 3.6 \micron, but the Circinus spectrum is very ``red'' and the $L$-band flux is very low. Instead, a fringe extraction using the $M$ band was necessary.

The data are then re-reduced using a custom python script\footnote{\href{https://github.com/jwisbell/matisse_lm_datareduction}{https://github.com/jwisbell/matisse\_lm\_datareduction}}. The custom pipeline uses the intermediate products of the DRS, specifically the complex cleaned interferograms (files called OBJ\_CORR\_FLUX). Using the 4.6 \micron~flux, the six fringes are identified and extracted in each exposure snapshot and each beam commuting device (BCD) configuration. Additionally, a bias ``fringe'' per frame is extracted far from the science fringes. The extracted fringes for each baseline and the extracted bispectra for each closure triplet are then bias corrected and temporally averaged incoherently (over the exposure cycle). More details are provided in Appendix \ref{app:datared}.

The resulting correlated fluxes and closure phases are computed for each BCD independently, and the final values are taken as the mean of the four BCD configurations. 
The final errors are the standard deviations of the four BCD configurations.  In the $L$ band the correlated flux errors on individual baselines are typically 1.8 mJy, and in the $M$ band the correlated flux errors are typically 5--10 mJy, in comparison to the typical $L$-band flux of 2--10 mJy and the typical $M$-band flux of 10--40 mJy. In both bands the closure phase errors are in general quite large, $\gtrsim 90^{\circ}$, and their use is limited.
This process was done for both the calibrators and Circinus, and the resulting observables were calibrated as usual. Reductions using the DRS and the custom pipeline for both a calibration star and for Circinus are shown in Fig. \ref{fig:circl_reduction} for comparison and validation of the approach.

The correlated flux, $F(u,v,\lambda)$ is then calibrated in the same way as in Paper I. 
The squared visibilities are finally calculated as $V^{2}(u,v,\lambda) = [F_{\rm targ}^{\rm cal}(u,v,\lambda) / F_{\rm targ}^{\rm tot}(\lambda)]^2$, where the total flux $F_{\rm targ}^{\rm tot}$ comes from the shortest baseline correlated fluxes; in this case we used the azimuthal maximum of the 30-35 m correlated fluxes as an estimate. The 30 m total flux and sample of all the correlated fluxes are shown in Fig. \ref{fig:cflux0} (the remaining correlated fluxes are shown in Appendix Fig. \ref{fig:cflux1}). We use this 30 m correlated flux rather than the ``zero-baseline'' correlated flux because squared visibilities can be scaled somewhat arbitrarily; the relative changes give substructure and the absolute changes are indicative of over-resolved emission. The 8.2 m total flux from ISAAC observations \citep[$F_{\rm nuc, L}=458.16\pm39.18$~mJy, $F_{\rm nuc, M}=676.41\pm44.58$~mJy; ][]{isbell2021} would result in extremely small squared visibilities ($\sim 3 \times 10^{-4}$) and cause numerical issues with little gain in understanding of the source.

\begin{figure*}
    \centering
    \includegraphics[width=0.95\textwidth]{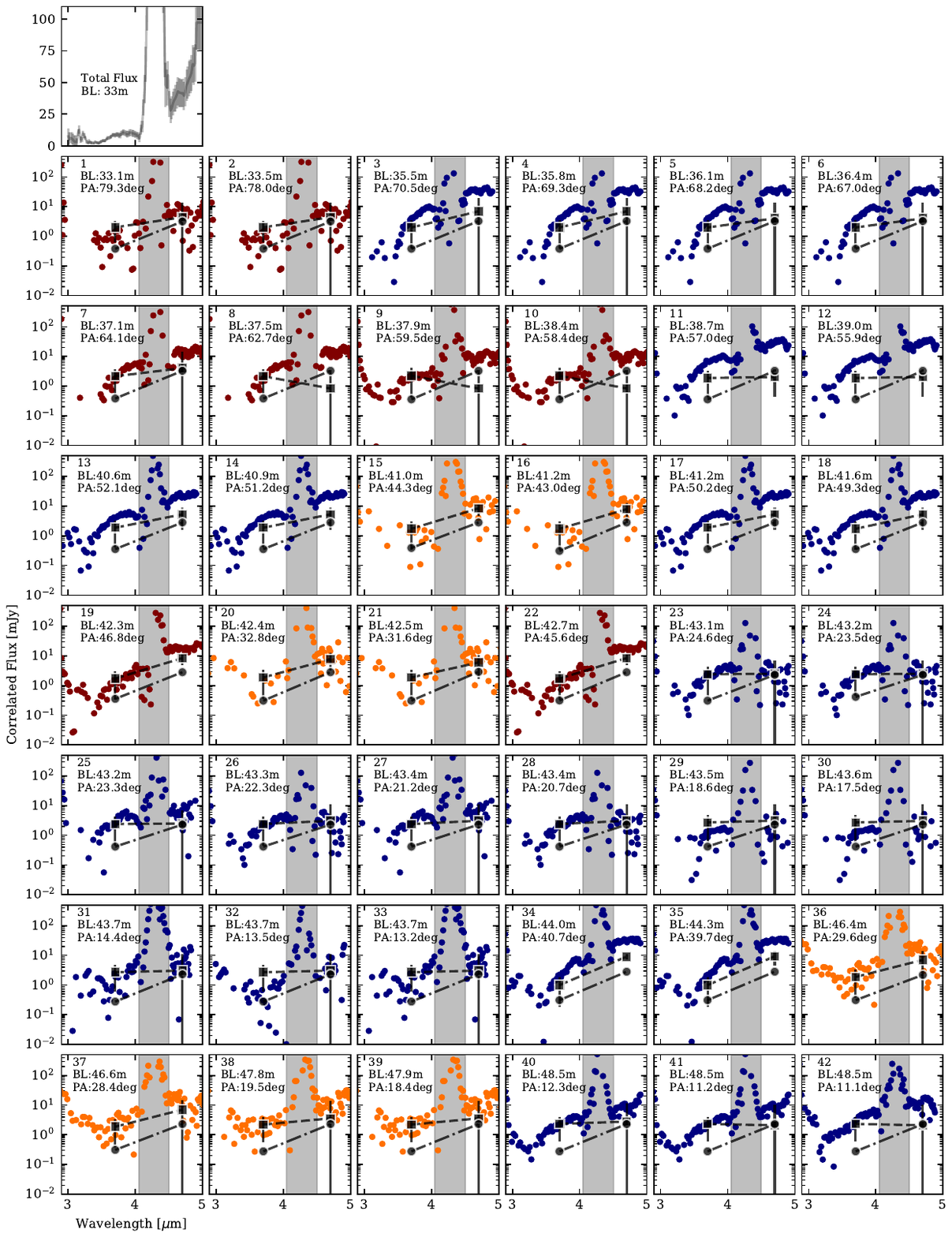}
    \caption[Excerpt sample of calibrated correlated fluxes for Circinus in the $LM$-band.]{{Excerpt sample of calibrated correlated fluxes for Circinus in the $LM$ band. The top-left panel shows the 33 m ``total flux'' used to compute squared visibilities. The colors indicate the observing date: March 2020 (blue), February 2021 (yellow), and May 2021 (red). Black squares and black circles are the predicted correlated fluxes from the image reconstructions and Gaussian model fits, respectively. Continued in Fig. \ref{fig:cflux1}.}}
    \label{fig:cflux0}
\end{figure*}

The closure phases are calibrated as in Paper I, but they instead use the results of the custom pipeline. Due to the low signal-to-noise ratio (S/N) for the visibilities on many baselines, only four closure phase triangles give reasonable values. The others are all dominated by noise, as mentioned above. The closure phases are shown in Appendix Fig. \ref{fig:t3phi0}.

\subsection{Single-dish observations}
This section describes single-dish observations of Circinus in the $L$ and $M$ bands from NACO and VISIR at the Very Large Telescope (VLT). These observations provide large-scale context for the interferometric images, linking the subparsec scales to the galactic scale. Additionally, we looked for signatures of flux variability.

\subsubsection{NACO observations}
We observed the Circinus nucleus with VLT/NACO in the Lp and Mp filters (3.4-4.2 \micron~ and 4.4-5.2 \micron, respectively) in burst mode in 2018 as part of program 0101.B-0446.
A point spread function (PSF) and flux calibrator, IRAS 14480-5828, was observed in concatenation. It is the
only target within 10$^{\circ}$ that has roughly similar color and brightness in the near-infrared and is
similar or brighter at $L$ and \textit{M} (\textit{W1}$\sim6$~mag, \textit{W2}$\sim 4.5$~mag). It is a basically unknown object, but is classified as a star in the SIMBAD database. It has no apparent optical counterpart.

We fit a two-dimensional Gaussian to each PSF-limited observation. This yields the full width at half maximum (FWHM) of the major and minor axes as well as the PA of the major axis for the target and for the calibrators \footnote{The fitted uncertainties on the FWHM and PA are $\lesssim 1\%$ and are omitted here for clarity.}.
The calibrator IRAS 14480-5828 seems unresolved in Mp with FWHM 0.13" x 0.13", PA=28$^{\circ}$. However, it turned out
to be extended in Lp with 0.20" x 0.19", PA=$112^{\circ}$).  From SED fitting of the \textit{Spitzer}/IRAC and WISE data, we obtain the following fluxes for IRAS 14480-5828 in Lp and Mp of 1501.6 mJy and 2313.0 mJy, respectively.
Because IRAS 14480-5828 is resolved in Lp, we used an alternative calibrator, 2MASS J16232835, with an Lp flux of
89.43 mJy. This star was observed as part of 0101.C-0924 (PI: Christiaens; unpublished)
in the same night as our Lp observation. The star is unresolved with 0.12" x 0.11" and
PA=167$^{\circ}$. 
The nucleus of Circinus appears to be extended with 0.17" x 0.14", PA=114$^{\circ}$, and 0.19" x 0.14",
PA=108$^{\circ}$ in Lp and Mp, respectively. The achieved angular resolution and depth is significantly better than in the archival NACO and ISAAC observations. 

The calibrated Lp and Mp fluxes for the Circinus nucleus are listed in Table \ref{tab:naco}.
Depending on the choice of the flux reference we obtain fluxes that differ by a factor
of $\sim2$. The likely reason for this is that at least one of our flux references is
variable. We see some evidence for this in the SED of IRAS 14480-5828 where
there are factor of~1.5 differences between the IRAC and WISE measurements. Flux calibrated images are shown in Fig. \ref{fig:naco_im} and fluxes extracted in 0.4" and 4.0" apertures are reported in Table \ref{tab:naco}. We adopted the values obtained by calibrating the $L$ band and $M$ band with 2MASS J16232835 and IRAS 14480-5828, respectively. 

The 0.4" Lp filter flux measurement presented herein is $\sim 50$~mJy larger than the $L$-band unresolved flux measured with ISAAC reported by \citet{isbell2021}. This difference is likely due to the use of aperture extraction in this work, while those authors did PSF extraction via Gaussian fitting of multiple components; their method attempted to separate resolved and unresolved flux. 

\begin{table}[]
    \centering
    \begin{tabular}{c|ccr}
         Filter & Calib. & Aperture ["] & Flux [mJy]   \\ \hline\hline
         Lp & IRAS 14480-5828 & 0.4 & $968 \pm 9.34$ \\
         Lp & IRAS 14480-5828 & 4.0 & $1044 \pm 56.3$ \\ \hline
         Lp & 2MASS J16232835 & 0.4 & $510 \pm 6.04$ \\
         Lp & 2MASS J16232835 & 4.0 & $685 \pm 60.4$ \\ \hline
         Mp & IRAS 14480-5828 & 0.4 & $1243 \pm 14.4$ \\
         Mp & IRAS 14480-5828 & 4.0 & $1315 \pm 137$ \\
    \end{tabular}
    \caption{Calibrated Circinus Lp and Mp fluxes from NACO observations taken in 2018.}
    \label{tab:naco}
\end{table}

\begin{table}[]
    \centering
    \begin{tabular}{c|ccr}
         Filter & Calib. & Aperture ["] & Flux [mJy]   \\\hline\hline
         M & HD 138538 & 0.4 & $\leq 2299$  \\
         M & HD 138538 & 4.0 & $\leq 2721$ \\ \hline
    \end{tabular}
    \caption{Calibrated Circinus \textit{M}-band fluxes from VISIR observations taken in 2017.}
    \label{tab:visir}
\end{table}

\subsubsection{VISIR M-band observations}
We observed the nucleus of Circinus with  VLT/VISIR in the \textit{M}-band filter in burst mode in 2017 as part of program 099.B-0235 (PI: Hoenig). During the observations, the VISIR \textit{M}-band filter suffered from a red leak, but we were able to roughly correct for this by subtracting a scaled PAH2\_2 filter image. Nevertheless, all the flux values reported here should be regarded as upper limits, including the typical 10\% systematic uncertainty on the flux of the calibrator. 

The nucleus of Circinus appears extended in both direct and PSF-subtracted images. The direct image is shown in Fig. \ref{fig:visir_im} and has a size of $0.33'' \times 0.30''$ with PA = 98$^\circ$, while the calibrator, HD138538, has a size of $0.22'' \times 0.18''$ with PA = 108$^\circ$, and was observed directly before Circinus in a similar direction of the sky. We measured the flux of Circinus using apertures of two different sizes: $0.4''$ and $4.0''$. The resulting fluxes are given in Table \ref{tab:visir}.

\subsubsection{Checking for variability}
\begin{figure}
    \centering
    \includegraphics[width=0.5\textwidth]{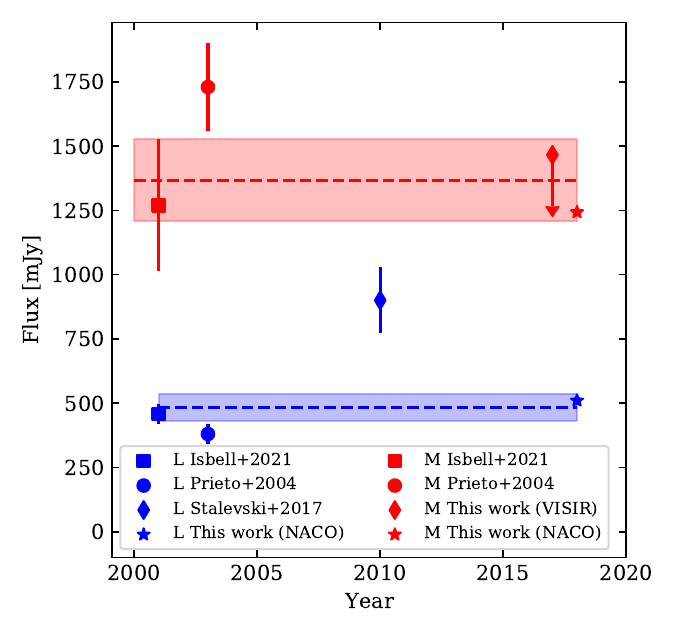}
    \caption{Circinus 0.4" fluxes in the $L$ and $M$ bands between 2000 and 2018. We compile the measurements of \citet{prieto2004}, \citet{stalevski2017}, \citet{isbell2021}, and this work. $L$-band values are plotted in blue, and $M$-band values are plotted in red. The mean flux for each band is plotted as a dashed line, with the standard deviation of the fluxes given as a shaded region. }
    \label{fig:variability}
\end{figure}

Because of the 0.4" aperture observations presented in this work, we can check the stability of the $LM$-band fluxes in Circinus over 18 years. We show the collected flux measurements in Fig. \ref{fig:variability}, bringing together data from \citet{prieto2004}, \citet{stalevski2017}, \citet{isbell2021}, and this work. We find that Circinus's $LM$ flux has remained fairly constant over the past decades. However, the $L$-band measurement from \citet{stalevski2017} is $\sim 2\times$ higher than the other three $L$-band fluxes; there are no apparent issues with the calibrator, HD129858 according to the MIR calibrator catalog of \citet{cruzalebes2019}. However, \citet{stalevski2017} used the MIDI observations of \citet{tristram2014} to measure in a 0.4" aperture, while \citet{isbell2021} uses single-dish observations in which the PSF was distorted or significantly larger than 0.4". The apparent difference could be due to this and/or the different method of measurement (PSF-fitting vs. aperture extraction).

\section{Modeling and image reconstruction}
\label{sec:imaging}
Both image reconstruction and model fitting rely on minimization of a cost function, $q$, which measures the flux distribution's similarity to a weighted combination of the closure phases and squared visibilities. 
As in Paper I and \citet{hofmann2016, hofmann2022}, we use a modified $\chi^2$ function: 
\begin{equation}
    q = \frac{\alpha}{N_{V^2}} \sum_{i=1}^{N_V} \frac{ (V^2_{\rm obs, i} - V^2_{\rm model, i})^2 }{\sigma_{\rm V^2, obs}^2 } + \frac{\beta}{N_{\phi}} \sum_{j=1}^{N_{\phi}} \frac{ (\phi_{\rm obs, j} - \phi_{\rm model, j})^2 }{\sigma_{\rm \phi, obs}^2 }.
    \label{eq:qrec}
\end{equation}
In the case of image reconstruction, a regularization term is added, yielding the cost function for an image, $I$, sampled at the vector of $uv$ coordinates, $\Vec{x}$,
\begin{eqnarray}
    J(I(\Vec{x})) = q(V_{\rm model}^2, \phi_{\rm model}) + \mu R(I(\Vec{x})),
\end{eqnarray}
where $q$ represents Eq. \ref{eq:qrec}, $V_{\rm model}^2$ and $\phi_{\rm model}$ are the squared visibilities and closure phases of the image flux distribution sampled at $\Vec{x}$, $\mu$ is the so-called ``hyperparameter'' that sets the amount of regularization applied, and $R$ the regularization function to be applied. 

We first attempted model-independent image reconstruction, as these are the first interferometric observations of Circinus in the $LM$ bands, and the source flux distribution was unknown. Following the relatively simple results of the image reconstruction, we performed follow-up modeling of the flux using elongated Gaussian components.

\subsection{Image reconstruction}
Following the procedure in Paper I, we use the package included in the DRS, Image Reconstruction Software using the Bispectrum \citep[IRBis;][]{hofmann2014,hofmann2022} to reconstruct images in the $LM$ band. We selected two wavelength bins in which to produce independent images: $3.7 \pm 0.1$ \micron~and $4.7 \pm 0.1$ \micron. Any spectral information within each bin is averaged, producing a series of ``gray'' images. Each bin is imaged with a range of regularization functions and hyperparameters (hereafter $\mu$; essentially a scaling on the amount of regularization), with the best selected via Eq. \ref{eq:qrec}.
We performed a grid search of the IRBis parameters, varying the field of view, the pixel number, the object mask scale, the regularization function, and the hyperparameter $\mu$. We used nonuniform weighting in the $uv$ plane, setting \code{weighting=0.5} in IRBis to de-emphasize the sparsely sampled and low-S/N points on baselines longer than $\sim 60$ m.

\begin{figure*}
    \centering
    \includegraphics[width=1\textwidth]{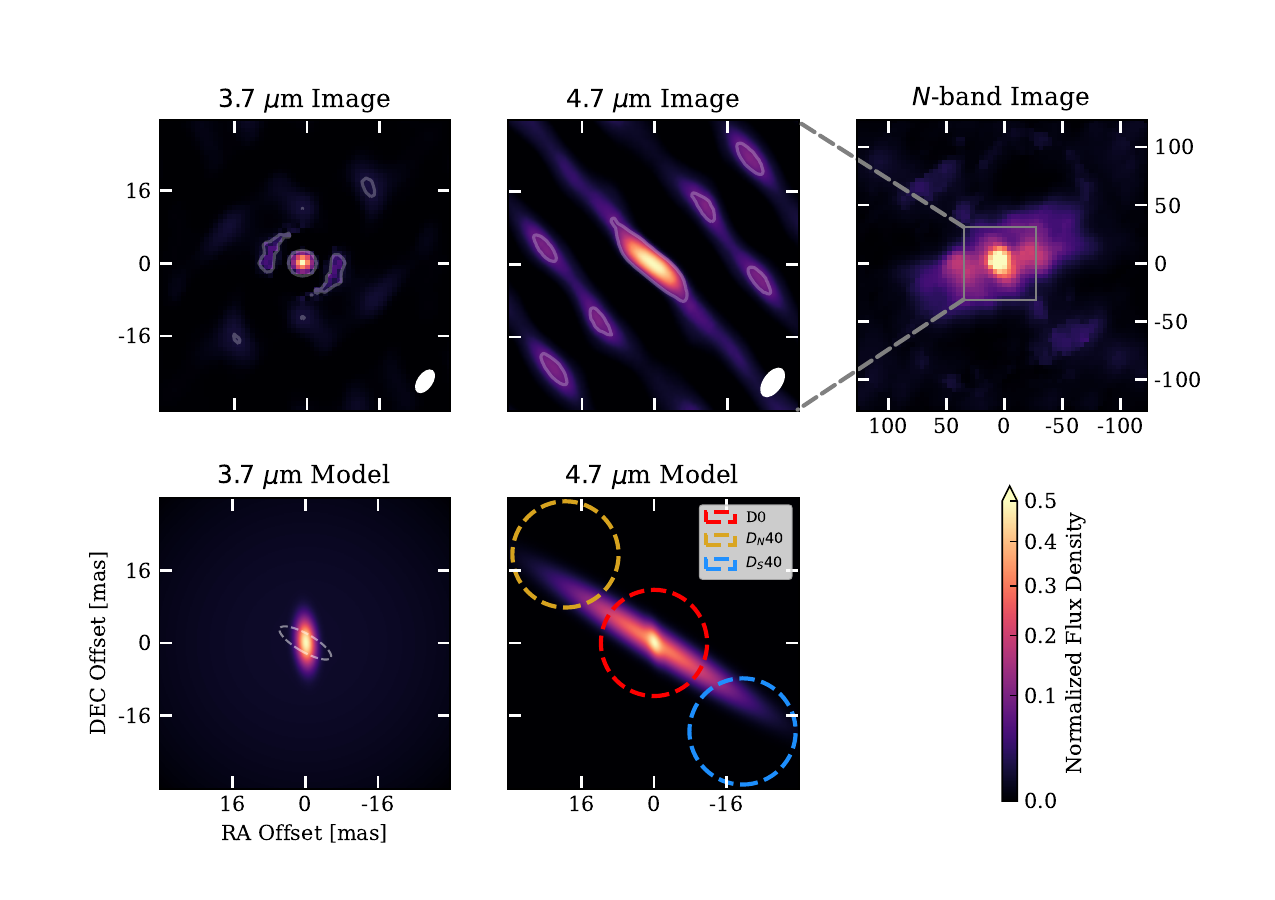}
    \caption[Images and models of the Circinus $LM$-band dust.]{Images and models of the Circinus $LM$-band dust. (\textit{top}) Image reconstructions at 3.7 and 4.7~$\mu$m. The contours are at $10\times \sigma_{\rm Im}$, estimated from the error maps produced by delete-$d$ jackknifing the $uv$ coverage. The FWHM of the estimated dirty beam is given in the bottom-right corner of each panel. Also included is the $N$-band continuum image from Paper I for reference. (\textit{bottom}) Gaussian model fits to the same wavelengths, specifically the 1+ model in the $L$ band and the two-Gaussian model in the $M$ band. The dashed ellipse in the $L$-band model image represents the 1$\sigma$ PA uncertainty of the fitted model.  The colored circles in the bottom-right panel illustrate the extraction apertures for the measured fluxes. }
    \label{fig:circl_immodel}
\end{figure*}

An initial best image is selected in each wavelength bin using minimization of Eq. \ref{eq:qrec}, 
and a follow-up round of imaging using the best regularization function and pixel scale is performed.
We give the final parameters for the reconstructions in Table \ref{tab:circl_imarec}. The images are shown in Fig. \ref{fig:circl_immodel} and the predicted correlated fluxes and closure phases from the reconstruction are overplotted in Figs. \ref{fig:cflux0} and \ref{fig:cflux1}-\ref{fig:t3phi3}; here it is apparent that the sparse $uv$ coverage and low S/N of the correlated fluxes have resulted in significant image artifacts, particularly in the $M$ band. We also see that the highest correlated flux values are not always matched in the $M$ band, usually when low correlated flux values are found at adjacent $uv$ points. Image errors are estimated as in Paper I, using delete-$d$ jackknifing. 

Despite the artifacts, the results of (1) a point source in the $L$ band and (2) a highly elongated component in the $M$ band at $\sim 45^{\circ}$ PA are robust above the noise (i.e., $> 10\sigma$, where $\sigma$ is estimated on a pixel-by-pixel basis using the delete-$d$ jackknifing method described in Paper I). The point-like source is present in the $L$ band, but it is not present in the $M$-band image reconstruction. There is, nonetheless, a peak in flux in the $M$-band image at the same pixel position as the point source. The disk-like component is roughly 4 mas in FWHM. This is comparable to the ideal resolution at 4.7 \micron~(3.7 mas), so the disk width is only marginally resolved. The structure extends 20.2 mas in FWHM along PA$\approx 46^{\circ}$. There are several secondary features that cross the $M$-band image along PA $\sim 40$; these are certainly artifacts due to their regular spacing, symmetry, and flux falloff with radius. Peaks along these lines indicate potential real polar flux, but they are at too low significance to analyze robustly.

\begin{table*}[ht!]
    \centering
    \caption{Final image reconstruction parameters.}
    \footnotesize
    \begin{tabular}{r|ccccccc}
        $\lambda$  & Reg.$^a$ & $\mu^b$ & FOV$^c$ & $N_{px}$  & Obj. Mask$^d$ & Cost$^e$& $\chi^2$$^{~g}$  \\
        $[\mu$m$]$  & Func.    &       & [mas] & - & [mas]       &  Func.  & [V$^2$,$\phi_{T3}$]  \\\hline \hline
        $3.7 \pm 0.1$   & 1 & 0.01 & 128& 128& 120 & 1 &[1.31,0.4]\\
        $4.7 \pm 0.1$   & 1 & 0.01  & 128 &256& 50 & 1 &[1.14,0.4]\\
    \end{tabular}
    \\
    \small{
    \textit{a}: the IRBis regularization function; \textit{b}: the weight on the regularization function (also known as the hyperparameter); \textit{c}: the field of view of the reconstructed image; \textit{d}: the radius of the object mask employed by IRBis in mas; \textit{e}: the cost function used in reconstruction, as described in Eq. \ref{eq:qrec} and in \citep{hofmann2022}; \textit{g}: the $\chi^2$ terms from the final images entering Eq. \ref{eq:qrec} for the squared visibilities and closure phases, respectively.}
    \label{tab:circl_imarec}
\end{table*}

\subsection{Gaussian modeling}
\label{sec:circl_gauss}
The image reconstructions result in simple components: a point source at 3.7 \micron~and a disk-like structure at 4.7 \micron. These simple structures can be modeled using elongated Gaussian components, which serves to minimize the effect of the dirty beam and image artifacts due to sparse $uv$ coverage and a small number of closure phases.

\begin{table*}[]
    \centering
    \caption{Fitted Gaussian parameters for the $LM$ bands.}
    \begin{tabular}{cc|cccc|c}
        $\lambda$ & $N_{\rm comp}$ &$\Theta$ & $r$ & $\phi$ & f & $\ln L$ \\
        $[\mu{\rm m}]$ & -  & $[$mas$]$ & - & $[$deg$]$ & - & - \\ \hline \hline 
        $3.7\pm 0.1$ & 1& $16.3_{-0.6}^{+0.8}$ & $0.63_{-0.05}^{+0.13}$ & $42.5_{-8.2}^{+7.4}$ & 1 & 257 \\\hline
        $3.7\pm 0.1$ & 1+& $6.0_{-1.5}^{+1.6}$ & $0.5_{-0.2}^{+0.2}$ & $14.3_{-9.3}^{+16.6}$ & 1 & 267 \\
         & & 128 & 1 & 0 & $0.07_{-0.01}^{+0.01}$ & \\\hline 
        $3.7\pm 0.1$ & 2& $6.5_{-1.7}^{+0.5}$ & $0.3_{-0.1}^{+0.2}$ & $3.3_{-1.2}^{+8.7}$ & 1 & 270 \\
        & & $18.9_{-0.6}^{+1.2}$ & $0.6_{-0.1}^{+0.2}$ & $44.0_{-8.2}^{+6.4}$ & $0.78_{-0.32}^{+0.15}$ & \\
        \\
        \\\hline
        $4.7\pm 0.1$ & 1&  $25.7_{-1.2}^{+2.4}$ & $0.2_{-0.1}^{+0.1}$ & $58.4_{-2.5}^{+2.5}$ & 1 & 247 \\\hline 
        $4.7\pm 0.1$ & 2& $30.2_{-1.3}^{+0.6}$ & $0.14_{-0.07}^{+0.10}$ & $57.6_{-3.1}^{+2.1}$ & 1 & $262$ \\
         &  &  $5.4_{-1.3}^{+0.6}$ & $0.4_{-0.2}^{+0.3}$ & $12.8_{-4.1}^{+10.7}$ & $0.6_{-0.2}^{+0.3}$ &  \\\hline
         $4.7\pm 0.1$ & 3&  $28.4_{-1.5}^{+2.2}$ & $0.13_{-0.09}^{+0.11}$ & $52.4_{-1.9}^{+3.6}$ & 1 & 263 \\
         &  & $3.7_{-0.3}^{+1.2}$ & $0.7_{-0.3}^{+0.2}$ & $79.5_{-4.1}^{+5.3}$ & $0.8_{-0.4}^{+0.1}$ &  \\
         &  & $48.8_{-1.4}^{+2.4}$ & $0.2_{-0.2}^{+0.4}$ & $84.4_{-5.4}^{+2.1}$ & $0.1_{-0.1}^{+0.2}$ &  \\
        
    \end{tabular}
    
    \label{tab:circl_gaussparams}
\end{table*}

The fitted Gaussian components are fixed to the center of the image and only their major axis $\Theta$, their minor-to-major axis ratio $r\equiv \theta/\Theta$, the PA of the major axis $\phi$, and their relative flux $f$ are allowed to vary. We fit a number of Gaussian components $N_{\rm comp}$ to each wavelength channel, but in general favor models with fewer components. The relative flux $f$ of one Gaussian is fixed to 1. The fitted Gaussians' parameters and error estimates are obtained through Markov-Chain Monte Carlo likelihood maximization. We sampled the parameter space using the package \pack{emcee} \citep{foreman-mackey2013}. The log-probability function to be maximized is given by the typical Bayesian formulation,
\begin{equation}
p(\vec{\theta}, c | \vec{x},y,\sigma) \propto p(\vec{\theta}) p(y | \vec{x}, \sigma, \vec{\theta}, c),
\end{equation}
with measurements $y = (V_{\rm model}^2, \phi_{\rm model})$ at $uv$ coordinates $\Vec{x}$, parameters $\vec{\theta}$, and error estimates $\sigma$ scaled by some constant $c$.
For maximum likelihood estimation, the log likelihood function for the models is written as
\begin{equation}
    \ln  L(y | \vec{x},\sigma,\vec{\theta} , c) = -\frac{1}{2} \sum_{n} \Big [ q(y) + \ln(2 \pi s_n^2) \Big ], 
\end{equation}
where $q$ (Eq. \ref{eq:qrec}) is the cost function for the squared visibilities and closure phases produced by the model with parameters $\Vec{\theta}$, $c$ represents the underestimation of the variance by some fractional amount, and $s_n^2= \sigma_n^2 + c^2f(x_n,\vec{\theta})^2 $. We estimate the best-fit value as the median of each marginalized posterior distribution and the $1\sigma$ errors from the values at $16^{\rm th}$ and $84^{\rm th}$ percentiles. 
Because the closure phases are very low S/N, we fix all components to the center of the image, and we set $\beta=0$ in Eq. \ref{eq:qrec} to fit the squared visibilities alone.

We first fit one Gaussian (i.e., $N_{\rm comp} = 1$) based on the simplicity of the image reconstruction, but Gaussian fits with more components were attempted as well. 
We favor models with fewer parameters based on the Akaike information criterion \citep[AIC;][]{akaike1981}. The AIC for a model with $k$ parameters and maximum likelihood $L$ is 
\begin{eqnarray}
    a = 2k - 2 \ln(L).
    \label{eq:circl_aic}
\end{eqnarray}
The model with the minimum value of $a$ is considered the ``best'' because it is a sufficient representation of the data without overfitting. For each fitted model with Gaussians fixed at the center, we fit up to $k = 4 N_{\rm comp} -1$ parameters (because the flux $f$ of one component is fixed).  The fit results for 3.7~$\mu$m with $N_{\rm comp} = 1$~and 4.7 \micron~with $N_{\rm comp} \in \{1,2,3\}$ are given in Table \ref{tab:circl_gaussparams} and the best fitting models are shown in Fig. \ref{fig:circl_immodel}.

In the $L$ band, a true single-Gaussian model does not perform well. It becomes large in order to produce low visibilities at short baselines, but then the long-baseline visibilities are far too low. An augmented single-Gaussian model was then fit, wherein a second, large component was added. This second component has fixed size ($128 \times 128$~mas) and orientation, but its flux is allowed to vary. It plays the role of over-resolved flux. This model (called 1+) gives a marked improvement in AIC over the single-Gaussian model with only one additional parameter. It results in a marginally extended source with FWHM $7.1 \times 2.8$ mas. A two-component Gaussian model produces a similarly extended source ($6.5 \times 2.0$ mas) with a much less extended second component. The two-component model produces the same AIC value as the 1+ model; for the rest of this work, we used the simpler 1+ model.

In the $M$ band, all three modeling results include a disk-like component with a similar size and orientation; this disk-like component reproduces the image reconstruction's morphology. In the two- and three-component fits, a point-like source is introduced on top of the disk. In the three-component model, a diffuse, roughly polar extended source is added. While its orientation is suggestive (based on the $N$-band polar emission), this additional component is disfavored by the AIC. If there is a signature of the polar dust in the $M$ band, it is at low significance, and more observations would be necessary to confirm it. The model with two Gaussian components is preferred at 4.7 $\mu$m, and it will be used for the rest of this work.  We note, however, that the flux in the central aperture (see the following section) differs by only $1\%$ between the one- and two-component models. The selected model is marginally resolved with a width of 4.2 mas (the 4.7 \micron~resolution is 3.7 mas) and has a major axis with FWHM $= 30.2_{-1.2}^{2.4}$ mas and PA$=57.6_{-3.1}^{+2.1 \circ}$. The increased size of the disk in the modeling when compared to the images likely compensates for the large-scale flux that is allowed in the imaging. This scenario is supported by the observation that the PA of the disk component decreases with an increasing number of fitted components.

The $M$-band model and image each show maximum extension nearly perpendicular to the beam rather than in the polar direction. This is not unprecedented when comparing to the $N$-band images of Paper I (and to some extent to the L-band image or model). In the $N$ band, while the maximum extension is in the polar direction, a significant amount of flux is perpendicular to the beam in a disk-like component. Similarly, the $L$-band model has uncertainties in PA that mean it could also be aligned with the $M$-band extension, and the $L$-band image shows secondary peaks along that same direction. We would argue that indeed all images and models in all bands either show or allow for flux in this direction. The apparent lack of polar dust in the $M$ band is most likely due to a combination of low dynamic range in image reconstructions and the fact that large-scale structures ($>$~16 mas $=$~0.3 pc) are resolved out by the interferometer.

\section{Measuring component temperatures}
\label{sec:bb}

In Paper I, circular apertures with diameter $23.4$ mas were used to extract the flux from each $N$-band image at a number of locations. We made the assumption here that the $LM$ models can be astrometrically matched with the $N$-band images by aligning the photocenters. This was done for NGC 1068 in \citet{gamezrosas2022}, and their cross-correlation matching done in the $N$ band was in the end equivalent to photocenter matching. Therefore, we used the same aperture diameter and distribution as Paper I. 

In the $L$ band, although there is a point source within the central aperture, D0, much of the flux is contained in the background component. This is true in both the imaging and in the 1+ model. Accordingly, only $86\%$ (in the Gaussian model) and $37\%$ (in the image) of the total $L$-band flux is contained in the aperture D0. 
In the $M$ band, the majority of the flux is found in the central aperture with only a minuscule amount falling in the disk apertures D$_{S}40$ and D$_{N}40$. In the imaging and modeling of neither band is there significantly measured flux in the polar direction. For these regions we present an upper limit from the ``sky'' background $F_{\rm upperlim.} \leq 2\sigma_{\rm sky}$ in the image reconstructions. All apertures from the $N$-band analysis that are not listed in Table \ref{tab:circl_exfluxes} are considered to have only upper limits: $F_{3.7\mu{\rm m}} \leq 1.2$~mJy and $F_{4.7\mu{\rm m}} \leq 2.2$~mJy. 

The Gaussian models and image reconstructions give slightly different morphologies, and therefore yield different flux measurements. We considered both sets of results independently. The 3.7 and 4.7 \micron~fluxes for the apertures from each imaging method are given in Table \ref{tab:circl_exfluxes}. The uncertainties in each case come from either the image or model uncertainty (each is described above) and the uncertainty on the total correlated flux (at 30 m). The total flux uncertainty ($\sim 5$~mJy at 4.7~$\mu$m) dominates in both imaging and modeling. 

\begin{table}[]
    \centering
    \small
    \caption{$L$- and $M$-band extracted fluxes.}
    \begin{tabular}{l|cc|cc}
        & Image & Reconstruction & Gaussian & Modeling$^a$ \\
        Aperture$^b$ & $F_{3.7\mu{\rm m}}$ & $F_{4.7\mu{\rm m}}$&$F_{3.7\mu{\rm m}}$ & $F_{4.7\mu{\rm m}}$ \\ 
         & $[$mJy$]$ & $[$mJy$]$ & $[$mJy$]$ & $[$mJy$]$ \\ \hline 
         D0 & $3.0\pm 1.7$ & $13.2 \pm 3.5$ & $7.0\pm 1.6$ & $28.5 \pm 7.8$ \\ 
         D$_S$40 & $0.4_{-0.4}^{+0.6}$ & $3.5\pm 1.1$& $\leq 2.2$& $3.5 \pm 1.1$ \\
         D$_N$40 & $0.4_{-0.4}^{+0.6}$ & $3.8\pm1.1$& $\leq 2.2$& $3.5 \pm 1.1$\\
    \end{tabular}
    \\
    \small{
    Error estimates contain the contribution from the total flux uncertainty as well as the model uncertainties. Image reconstruction fluxes are typically lower than the Gaussian models because they include background flux and artifacts, which both take away flux from the primary components.
    \textit{a}: $L$-band fluxes come from the 1+ model and $M$-band fluxes from the two-component model, but both the one-component and two-component Gaussian models give similar (within 1\%) extracted fluxes.
    \textit{b}: Apertures from the $N$-band analysis that are not listed can be considered to be upper limits ($\leq 2\sigma_{\rm im}$), with $F_{3.7\mu{\rm m}} \leq 1.2$~mJy and $F_{4.7\mu{\rm m}} \leq 2.2$~mJy.
    }
    \label{tab:circl_exfluxes}
\end{table}

\subsection{Blackbody fitting}

We fit a two-blackbody (BB) curve with absorption to each aperture-extracted spectrum with the form
\begin{equation}
    I(\lambda, T, A_v) = \sum_{i=1}^2 \eta_i BB_{\nu}(\lambda,T_i) e^{ \frac{-A_{V,i}} { 1.09~\tau(\lambda)/\tau_v} }  ,
    \label{eq:bb}
\end{equation}
where $\eta$ is an absolute flux scaling due to the filling factor of the dust in the aperture, $\tau(\lambda) / \tau_v =\kappa(\lambda) / \kappa_v $, and we use the standard interstellar medium $\kappa(\lambda)$ profile from \citet{schartmann2005}, which is based on the standard interstellar medium profile of \citet{mathis1977}. 

Fitting of $T$, $\eta$, and A$_{\rm V}$ to the \textit{LMN} SEDs is done in two iterations using Markov chain Monte Carlo sampling with the package \pack{emcee} \citep[][]{foreman-mackey2013}, similar to the approach in Sect. \ref{sec:circl_gauss}. Final values in each iteration are the median of the marginalized posterior probability distribution. The 16th and 84th percentiles of the resulting temperature and extinction distributions are used as the $1\sigma$ fit uncertainties. 

For the first BB component, we use uniform prior probability distributions with $T_1 \in (100,500]$ K, $\eta_1 = 1$, and A$_{\rm V,1} \in [20, 37]$ mag. These priors are based on the $N$-band fit results, particularly $A_{\rm V} = 28.5_{-7.7}^{+8.5}$~mag for D0 (see Paper I). The second component is forced to be strictly hotter and smaller than the first component, resulting in the uniform priors $T_2 \in (500,1500]$~K, $\eta_2 \in (10^{-3},0.1]$, and $A_{V,2} \in (0,700]$~mag. 
For the central aperture, D0, which should cover the sublimation zone, another set of priors is also used. They come from the assumption that dust is the sublimation temperature is indeed present but can be heavily obscured. We estimated a representative sublimation radius for silicate dust at 1500 K using the formula from \citet{barvainis1987}:
\begin{equation}
\label{eq:barvainis}
r_{\rm sub} = 1.3 L_{UV,46}^{0.5} T_{1500}^{-2.8}~\rm{pc}.
\end{equation}
As \citet{moorwood1996} report $L_{UV} = 5\times 10^9 L_{\odot}$, we estimate $r_{\rm sub} = 0.05$~pc = 2.8 mas at a distance of 4.2 Mpc. This gives an upper limit on $\eta_2 \leq (2.8 / 11.7)^2 = 0.06$, where $11.7$~mas is the aperture radius. Therefore, we define the priors of the second component: $T_2  = 1500$~K, $\eta_2 \leq 0.06$, and $A_{V,2} \in (0,700]$~mag. This fit gives a rough estimate on the minimum amount of extinction necessary to hide dust at the sublimation temperature. It is important, however, to note that for different dust compositions and grain sizes, the sublimation radius can vary from 0.05 to 0.2 pc. This is compounded by uncertainties in the luminosity of the AGN. The value we adopted is meant only as a fiducial, representative value that is roughly the mean radius of the sublimation zone. 

\begin{figure*}
    \centering
    \includegraphics[width=0.44\textwidth]{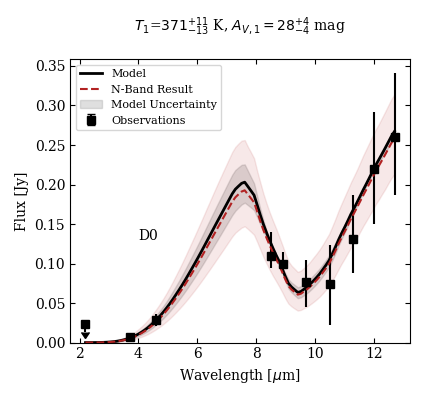}
    \includegraphics[width=0.44\textwidth]{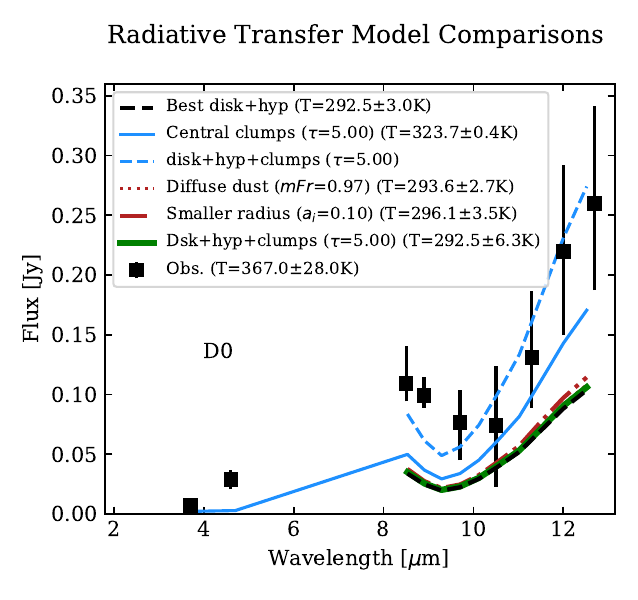}  \\
    \includegraphics[width=0.44\textwidth]{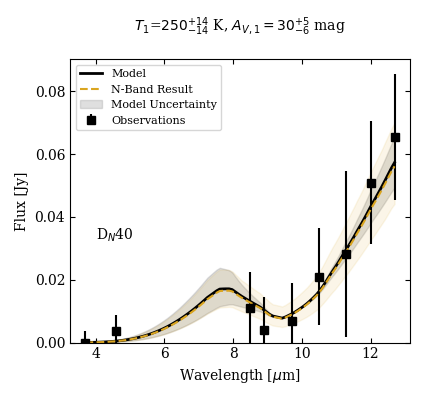}
    \includegraphics[width=0.44\textwidth]{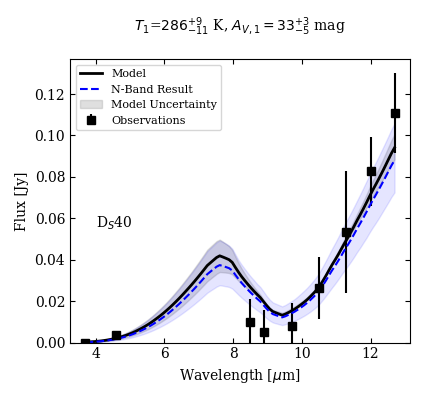}
    
    \caption[One-blackbody fits for the aperture-extracted Circinus \textit{LMN} fluxes.]{One-BB fits for the aperture-extracted Circinus \textit{LMN} fluxes. The colors are the same as in Fig. \ref{fig:circl_immodel}, with D0 in red, D$_N$40 in yellow, and D$_S$40 in red. The top-right plot compares the D0 fluxes to the various RT model modifications. 
    The fits using the $N$-band data alone are included for comparison. In aperture D0 the $K$-band measurement from \citet{burtscher2015} is included as an upper limit for the near-infrared flux. }
    \label{fig:circl_sedfit_1}
\end{figure*}

The recovered temperatures for the one- and two-component BB fits to the image reconstruction fluxes and to the Gaussian model fluxes are given in Table \ref{tab:circl_imarecsed}. The fitted SEDs for the Gaussian models using one and two BB fits are shown in Figs. \ref{fig:circl_sedfit_1} and \ref{fig:circl_sedfit_2}, respectively. The nuclear $K$-band flux from \citet{burtscher2015} for Circinus is shown in Figs. \ref{fig:circl_sedfit_1} and \ref{fig:circl_sedfit_2}, and it serves as an upper limit on the near-infrared flux; all extrapolated BB flux values are far below this limit. 

For the aperture D0, the image reconstructions' fluxes result in lower fitted temperatures than the Gaussians'. However, the fitted $T_1$ values in both the images and the Gaussian models are consistent within the uncertainties to the temperature inferred from the $N$ band alone ($367_{-26}^{+30}$~K). 
The fitted ``cool'' component temperature $T_1$ has essentially the same value both with and without the additional hot BB.
Importantly, in neither the images nor the models is an additional, hot component necessary. Large extinction values ($\sim 450$~mag) are preferred, and the hot component makes up only 0.2\% of the 3.7 $\mu$m flux and 1.3\% of the 4.7 $\mu$m flux. In fact, looking at the posterior probably distributions for all apertures, only second component extinction values $\gtrsim 250$~mag are allowed in any of the fits. 

For the disk apertures, D$_N$40 and D$_S$40, the BB-fit results are consistent from both the modeling and image reconstruction fluxes. The results, furthermore, are consistent with the $N$-band fitted temperatures and extinctions. A second component is once again disfavored due to the large fitted extinction values and the fact that the fit is not markedly improved with the addition of this component.

\begin{table*}[]
    \centering
    \caption{Temperature fit results.}
    \begin{tabular}{cc|ccccc}
        Image Reconstruction &&&&&& \\ \hline
        Aperture & $N_{\rm BB}$  &$T_1$ & A$_{V,1}$ & $\eta_2$& $T_2$  & A$_{V,2}$  \\
            & - &   $[$K$]$ &  $[$mag$]$ & -&$[$K$]$  & $[$mag$]$  \\ \hline 
        D0  &1&$343_{-5}^{+6}$ & $22.7_{-1.4}^{+2.4}$ &-&-&- \\ 
        D0 & 2 & $343_{-5}^{+6}$ & $22.6_{-1.2}^{2.2}$ & $0.05_{-0.03}^{+0.04}$ & $831_{-306}^{+407}$ & $479_{-179}^{+146}$\\
        D0 & 2* & $343_{-5}^{+6}$ & $22.6_{-1.3}^{2.4}$ & $0.03_{-0.02}^{+0.02}$ & 1500* & $520_{-135}^{+125}$\\ \hline
        D$_{N}40$ & 1 & $266_{-16}^{+13}$&$32.5_{-5.5}^{+2.7}$& -& -& \\
        D$_{N}40$ & 2 & $252_{-16}^{+15}$&$29.9_{-5.7}^{+4.5}$& $0.06_{-0.03}^{+0.03}$&$1042_{-371}^{+314}$ & $340_{-91}^{+74}$ \\ \hline 
        D$_{S}40$ & 1 & $286_{-11}^{+8}$&$33.0_{-4.3}^{+2.3}$& -& -& \\
        D$_{S}40$ & 2 & $281_{-13}^{+10}$&$32.3_{-5.3}^{+3.0}$& $0.05_{-0.03}^{+0.04}$&$840_{-329}^{+434}$ & $403_{-165}^{+184}$ 
        \\\\
        
        Gaussian Models &&&&&& \\ \hline
        \hline
        Aperture & $N_{\rm BB}$  &$T_1$ & A$_{V,1}$ & $\eta_2$& $T_2$   & A$_{V,2}$  \\
            & - &   $[$K$]$ &  $[$mag$]$ & -&$[$K$]$ &  $[$mag$]$  - \\ \hline 
        D0  &1 & $370_{-12}^{+11}$& $28.4_{-3.7}^{+3.9}$  &-&-&- \\ 
        D0  &2 & $367_{-15}^{+13}$  &$27.8_{-4.2}^{+4.2}$ & $0.05_{-0.03}^{+0.03}$&$891_{-354}^{+399}$ &$413_{-191}^{+197}$ \\
        D0  &2*& $367_{-15}^{+13}$ & $27.7_{-4.2}^{+4.0}$ &$0.03_{-0.02}^{+0.02}$& 1500* & $459_{-178}^{+164}$ \\ \hline 
        D$_{N}40$ & 1 & $250_{-15}^{+15}$&$29.7_{-5.7}^{+4.7}$& -& -& \\
        D$_{N}40$ & 2 & $250_{-14}^{+13}$&$30.0_{-5.6}^{+4.3}$& $0.05_{-0.03}^{+0.03}$&$880_{-335}^{+416}$ & $433_{-178}^{+183}$ \\ \hline 
        D$_{S}40$ & 1 & $301_{-13}^{+10}$&$42.7_{-7.0}^{+5.6}$& -& -& \\
        D$_{S}40$ & 2 & $281_{-13}^{+10}$&$32.1_{-5.3}^{+3.1}$& $0.05_{-0.03}^{+0.04}$&$904_{-362}^{+401}$ & $421_{-148}^{+173}$ 
    \end{tabular}
    \label{tab:circl_imarecsed}
\end{table*}

\begin{figure}
    \centering
    \includegraphics[width=0.5\textwidth]{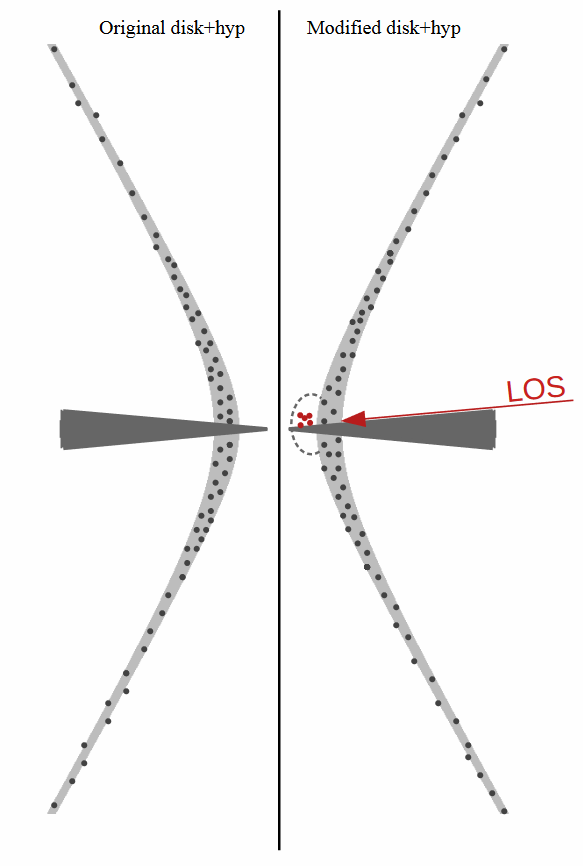}
    \caption{Schematic of the dust within the RT models that best reproduce the observed dust temperatures. The basis of the model comes from \citet{stalevski2019}, and a rough parameter range is given in Paper I. Here we have added a cluster of dust clouds above the disk and behind the hyperbolic cone, at a position that in projection corresponds to the central aperture. These clouds could represent, for example, a puffed-up sublimation zone, freshly launched winds, or a smoother boundary for the hyperbolic cone. The spatial resolution of these observations is not sufficient to distinguish between these scenarios. } 
    \label{fig:schematic}
\end{figure}

\subsection{Comparisons to radiative transfer models}
To describe the circumnuclear dust, a disk+wind model has become the consensus in both RT modeling \citep[e.g.,][]{stalevski2017,stalevski2019,honig2017} and hydrodynamical modeling \citep{williamson2020}. 
Specifically, \citet{stalevski2017,stalevski2019} undertook RT  modeling of VISIR imaging data, the MIR SED, and MIDI interferometric data of Circinus. Their best-fitting model \citep[presented in][]{stalevski2019} consists of a compact, dusty disk and a hollow hyperbolic cone extending in the polar direction (hereafter disk+hyp). In this modeling, a parameter grid for the RT models was searched such that the overall SED as well as the interferometric observables were well reproduced. This was not a model fit, but rather an exploration of the parameter space. This model was shown in Paper I to broadly agree with the $N$-band MATISSE data, though new constraints on the clump number density were reported. 

In this section we explore the applicability of those RT models and their SEDs to the combined \textit{LMN}-band results.  We used the same procedure for the RT model images as for the observations: we extracted fluxes in each of the 23.4 mas apertures and fit temperatures and extinctions using Eq. \ref{eq:bb} to the disk+hyp model grid at $\lambda \in [3.7, 4.7, 8.5, 8.9, 9.3, 9.7, 10.1, 10.6, 11.0, 11.5, 12.0, 12.5]$ \micron. The total flux of each RT model image is scaled such that the central $500 \times 500$ mas flux matches the total MATISSE-observed flux at that wavelength.
 
The most striking difference between the RT models and the observations is found in the central aperture. Both the extracted flux and the fitted temperature are found to be low in the disk+hyp models compared to the observations ($370_{-12}^{+11}$ K vs. $290$~K). We modified the disk+hyp models in several different ways to determine where additional warm dust could plausibly be found: 

\begin{enumerate}
    \item The inner radius of the fiducial disk+hyp model is 0.16 pc. We decreased this to 0.1, which increases the flux impinging the inner hyperboloid walls. This inner radius change produced only a small difference in fitted temperature ($\sim 4$ K)
    \item Diffuse dust with mass fractions of (0.7 to 0.97) was distributed between the clouds of the fiducial disk+hyp model. Diffuse dust actually decreased the fitted temperature in some cases, likely due to more self-shielding of the dust structures. 
    \item Finally, we added a cluster of clouds positioned just above the disk and inside the hyperboloid at a distance of 0.15 pc ($\sim 3 r_{\rm sub}$). This distance was set by the temperature versus radius estimates of Eq. \ref{eq:barvainis}; much closer in and the clouds would both be too hot (we see no evidence of $T\sim1000$~K dust) and/or blocked by the disk. In a standalone model, they were able to reach temperatures $\sim 330$K. When added to the existing disk+hyp model, this effect is reduced, but changing the optical depth or clump density of the cone wall would make it possible to glimpse these clouds more directly. 
\end{enumerate}

In each case, we measure the temperature in the central aperture, D0. We show the predicted temperatures and SEDs in Fig. \ref{fig:circl_sedfit_1}. We find that neither decreasing the inner radius nor adding diffuse dust made significant changes to the measured temperature. On the other hand, adding clouds above the disk but inside the hyperboloid greatly increased the measured temperatures, bringing the models into much better agreement with the observations. We show a schematic of the resulting RT model in Fig. \ref{fig:schematic}. We did not constrain the amount or exact location of these clouds because they are found on scales that remain spatially unresolved in our data. We instead emphasize that this simple modification of the nominal disk+hyp model at the base of the cone causes a large qualitative improvement in the agreement between the (modified) disk+hyp models and the observations.

\section{Discussion}
\label{sec:discussion}
Both the image reconstructions and the Gaussian models of the $LM$-band data are remarkably consistent with the structures imaged in the $N$ band. Specifically, in the $M$ band we recover a thin disk at PA$=57.6_{-3.1}^{+2.1 \circ}$. This disk is $4.2_{-2.2}^{+3.2}$~mas FWHM across -- comparable to the ideal 3.73 mas MATISSE resolution at 4.7~\micron~(note that the actual resolution is likely lower due to sparse and elongated \uv\  coverage). The $M$-band disk width is thus at best marginally resolved, and it is therefore consistent with the edge-on measurements of Paper I and the general trend of increasing inclination toward the center of Circinus \citep{izumi2018}. Additionally, in the $M$-band image reconstruction (Fig. \ref{fig:circl_immodel}), we see relatively bright artifacts in the polar direction. This suggests the presence of over-resolved flux at these scales, which we also see in single-dish ISAAC and NACO images at $\sim 400$ mas resolution \citep[this work and][]{isbell2021}.

Taken together, the single-dish and interferometric data suggest that the observed $LM$ emission relates more to large-scale dust structures than to the scales or dust temperatures observed with dust-reverberation mapping \citep[e.g.,][]{koshida2014}. We see little evidence of temporal changes in the single-dish flux over the last decades (Fig. \ref{fig:variability}). This lack of variability indicates that the emission originates not in directly illuminated, fast-moving clouds near the AD, but rather from a more distributed, large-scale dust structure. This conclusion is supported by the interferometric images and SED fits. We show that the \textit{LMN} emission can be simply explained by a single moderately warm dust structure missing any direct sightlines to very hot dust clouds. Specifically, aperture D0 is dominated by $\sim 367$~K dust emission, and any hotter dust emission is shown to be hidden by $A_{\rm V}\gtrsim 250$~mag of extinction. Furthermore, the morphological continuity between the \textit{LMN} bands suggests that the emission traces the same structure from $\sim 0.1$~pc to $> 1$~pc scale. 

It is tempting to conclude that the $M$-band image reveals a warp in the central disk. There is much evidence of a tilted (with respect to the ionization cone) AD in Circinus, for example the warped H$_2$O maser emission \citep[][]{greenhill2003}, fits of RT models to MIDI data \citep{stalevski2019}, and recent polarization mapping data \citep{stalevski2023}. In short, it seems that the warped maser emission does indeed trace a warped or tilted AD, which asymmetrically illuminates the circumnuclear dust. This was a possible explanation of the flux- and temperature-enhanced regions of the polar dust imaged in the $N$ band in Paper I. Radiation transfer modeling of dust around a warped maser disk by \citet{jud2017} showed that the warp would indeed be visible in the warm, innermost circumnuclear dust. Despite the evidence of a warp from other observations, we caution that the $M$-band closure phases used to reconstruct our image have very large ($>100^{\circ}$) errors in most closure triangles. So, while the correlated fluxes strongly constrain the spatial scale of the emitting $M$-band flux, the specific morphology relies on a handful of significant measurements. Either repeated observations under excellent conditions or improvements in data processing would be necessary to confirm the presence of a true warp in the $M$-band-emitting dust.

Both the image reconstructions and Gaussian models give the same SED and fitted temperature results. We extracted the flux in three apertures with diameters of 23.4 mas. These are a subset of the 13 apertures studied in Paper I, allowing us to study the spatially resolved \textit{LMN}-band SEDs. In all apertures, a single BB temperature is sufficient to describe the measured fluxes. There is no compelling indication of the expected hot, approximately sublimation temperature dust in the central aperture. This begs the question whether it is simply not there, or whether it is present and being obscured by the disk component. If the hot dust is simply not there, this raises serious questions and is difficult to reconcile with the latest radiation transfer and hydrodynamical modeling \citep[e.g.,][]{wada2016}. When we assume the dust is there, we find that very large extinction values ($\gtrsim 250$~mag) are necessary to match the observed flux in the central aperture. This value, however, is not incredible. Using the X-ray hydrogen column density to extinction relation of \citet{predehl1995}, 
\begin{equation}
    N_H~[{\rm cm}^{-2} / A_{\rm V}] = (1.79 \pm 0.03) \times 10^{21},
\end{equation} 
and the value N$_{\rm H} = 10^{24}~{\rm cm}^{-2}$ from \citet{matt1999}, we find ${\rm A}_{\rm V} = 558.6$~mag. Recently, ALMA observations by \citet{izumi2023} indicate extinction toward the AD of $A_{\rm V} = 210-440$~mag. These estimates are fully in line with our fitted value. Our fitted value is also in broad agreement with predictions from the disk+hyp RT model; the $A_{\rm V}$ of the model at the given inclination, averaged over the azimuthal angle, is approx $A_{\rm V} = 211$~mag \citep{stalevski2019}; however, this might be a lower limit, since the model grid did not include optical depth values larger than $\tau_{9.7}=15$ for the disk. All in all, it seems that the optically thick, geometrically thin disk plays the role of the dusty obscurer in the classic dusty torus model.

We test several ad hoc modifications to the disk+hyp models of \citet{stalevski2017, stalevski2019} in order to better reproduce the observed fluxes and temperatures of the central aperture, D0. We find that a cluster of clouds positioned above the disk and inside of the hyperboloid at a distance of 0.15 pc ($\sim 3 r_{\rm sub}$) matches the observations most closely. While we did not constrain the exact scale or origin of these clouds, they indicate a thicker hyperboloid wall in this region and/or a puffed-up disk near the center. In both cases, they indicate (dynamic) dust structures in this region strongly affected by the asymmetric radiation pressure of the tilted AD \citep[see, e.g., Paper I][for evidence of this tilt]{stalevski2019, stalevski2023}. A puffed-up inner region has been previously proposed by \citet{honig2012}. Within a zone of 5-10 sublimation radii (0.25-0.5 pc or 12-24 mas), the predicted puffed-up region described in \citet{honig2019} (based on 3-5 $\mu$m SEDs) should be resolved with our MATISSE observations. As we imaged this inner region for the first time, we placed constraints on the scale of this puffed-up zone in Circinus, roughly 3 $r_{\rm sub}$. If the puffed-up region indeed extends farther (to $\sim 10~r_{\rm sub}$), it could contribute to the ${\rm A}_{\rm V} \sim 500$~mag obscuration of the ``missing hot dust'' mentioned above. Future high-resolution MIR observations of other Seyfert galaxies can explore how typical the circumnuclear dust in Circinus is. Additionally, the dust in this regime could be related to the radiation-driven fountain suggested in \cite{wada2012}. Whether the new clumps represent failed flows or freshly launched dusty outflows is unclear. The latter would represent the onset of feedback with the host galaxy. Future, very high spatial resolution infrared spectra in this region could give an idea of the dust dynamics in Circinus.

\section{Conclusion}
\label{sec:conc}
Following up on the $N$-band analysis of Paper I, we present the first-ever $L$- and $M$-band interferometric observations of Circinus. These observations allowed us to reconstruct images and fit Gaussian models to the $L$- and $M$-band data. Using these images and models:
\begin{enumerate}
    \item We find a thin disk whose width is marginally resolved (0.08 pc = 4.23 mas). This disk is shown to be the spectral continuation of the disk imaged in the $N$ band to shorter wavelengths, as the measured fluxes correspond to the fitted $N$-band temperatures. In addition to this thin disk, there is a point-like source found in the $L$ and $M$ bands that was identified with the $N$-band point source based on the $LMN$-band SED fit. 
    \item We show that there is no trace of hot dust ($T \sim 1500$ K) in the circumnuclear dust structure of Circinus. By assuming the dust is there, we find that obscuration within the disk of A$_{\rm V} \gtrsim 250$~mag is necessary to reproduce the measured fluxes. With dust extinction this high, the imaged disk could then play the role of the obscuring ``torus'' in the unified scheme of AGNs.
\end{enumerate}

We also explore the parameter space of the disk+hyp models from \citet{stalevski2019}, identifying a simple modification that better matches the observations toward the center. We added a cluster of dust clouds above the disk but inside the radius of the base of the hyperbolic cone, at a position that in projection corresponds to the central aperture. These clouds could represent, for example, a puffed-up sublimation zone, freshly launched winds, or a smoother, radially wider boundary for the base of the hyperbolic cone. The spatial resolution of the presented observations is not sufficient to distinguish between these scenarios, but the presence of these clouds provides a much better match to the observed temperature distribution of the circumnuclear dust in Circinus than the standard disk+hyp models. In the future, we plan to test the applicability of this inner dust structure to other MATISSE-observed AGNs.

\begin{acknowledgements}
The authors would like to thank the late Prof. Dr. Klaus Meisenheimer for his valuable contributions to both this work and to their growth as scientists, through his inspiring supervision, thoughtful questions, and ambitious vision for the VLTI.
We thank the VLTI Paranal staff for all their help and good company during observations.
M.S. is supported by the Ministry of Science, Technological Development and Innovations of the Republic of Serbia through the contract No.~451-03-9/2023-14/200002.
This research has made extensive use of NASA’s Astrophysics Data System; the SIMBAD database and VizieR catalogue access tool, operated at CDS, Strasbourg, France; and the python packages \pack{astropy}, \pack{emcee}, \pack{scipy}, and \pack{matplotlib}.
The authors thank the anonymous referee for their helpful comments which greatly improved this paper.
\end{acknowledgements}

%
\bibliographystyle{aa} 
\bibliography{circinusL} 
%

\begin{appendix}
\section{Details on custom data reduction}
\label{app:datared}
The MATISSE DRS has been optimized for sources with an $L$-band flux $F_{3.6} > 0.5$~Jy or an $N$-band flux $F_{12} \gtrsim 1$~Jy. Moreover, it uses 3.6~$\mu$m as the reference wavelength for the estimation of the optical path delay (OPD) in the $L$ band. This works well for bright sources or for blue sources (such as young stellar objects). For red objects, such as Seyfert 2 galaxies, OPD corrections in the $L$ band are severely impacted by low S/N and so the $M$ band should be used. At the time of writing, there is no option to change the OPD reference wavelength. For the data reduction of Circinus, it was necessary to manually extract the fringes using the $M$ band as a reference. 

Fortunately, the Charge-Coupled Device (CCD) calibration steps and the production of the clean interferogram could be taken directly from the DRS. These steps resulted in a time series of two-dimensional Fourier transforms of the interferogram (called \code{OBJ\_CORR\_FLUX\_*.fits}). In each frame, the positions of the fringes could be measured and shifted (in other words, phase-corrected) before the correlated flux and closure phase could be extracted using the following methods; the fundamental equations we used can be found in the MATISSE instrument paper, \citet{lopez2022}.

\subsection{Squared (incoherent) correlated flux} The squared correlated flux\footnote{The terms correlated flux and coherent flux are used interchangeably throughout this work.} $C_{ij}^2$ for a baseline $B_{ij}$ is given as
\begin{eqnarray}
    C_{ij}^2(\lambda) = \sum_{\Vec{u}} \langle |I(\Vec{u},\lambda, t)|^2 - \beta \rangle _{t},
    \label{eq:matisse_incoh_corrflux}
\end{eqnarray}
where $\langle ... \rangle_t$ is the time average, $\Vec{u}$ is the spatial frequency integrated between $(\Vec{B}_{ij}-D)/\lambda$ and $(\Vec{B}_{ij}+D)/\lambda$, $D$ is the pupil diameter, and $\beta$ is the bias present in each fringe peak (estimated from the value between fringe peaks). The pupil here refers to the spatial-filter pinhole with diameter $1.5\lambda/d$ in the \textit{L} band and $2\lambda/d$ in the $N$ band with $d$ representing the telescope diameter. 

\subsection{(Coherent) correlated flux} The correlated flux is integrated coherently in the form
\begin{eqnarray}
    |C_{ij}(\lambda)| =  \Big | \sum_{\Vec{u}} \langle I(\Vec{u},\lambda, t) e^{-i \phi_{\rm atm}(\lambda, t)} \rangle_t \Big|,
    \label{eq:matisse_coher_cflux}
\end{eqnarray}
where once again $u$ is the spatial frequency integrated between $(\Vec{B}_{ij}-D)/\lambda$ and $(\Vec{B}_{ij}+D)/\lambda$ and $D$ is the pupil diameter.

\subsection{Closure phase}
Recalling that interferometric observations are complex numbers that have both an amplitude and a phase (the argument of the complex number), one can compute the closure phase for a telescope triplet $ijk$, $\phi_{ijk}$ in the following way:
\begin{eqnarray}
    \phi_{ikj}(\lambda) = Arg \Big[\sum_{\Vec{u}_1,\Vec{u}_2} \langle I(\Vec{u}_1,\lambda, t) I(\Vec{u}_2,\lambda, t) I^*(\Vec{u}_1+\Vec{u}_2, \lambda, t) \rangle_t  - \gamma  \Big].
\end{eqnarray}
The coordinates $\Vec{u}_1$, $\Vec{u}_2$, and $\Vec{u}_1+\Vec{u}_2$  correspond to the baselines $\Vec{B}_{ij}$, $\Vec{B}_{jk}$, and $\Vec{B}_{ik}$, respectively, integrated over the pupil width as for the squared correlated flux. The parameter $\gamma$ represents the photon bias in the bispectrum. This photon bias contains both an additive and a multiplicative term due to the combination of the fringes. The estimation of this bias is nontrivial, and the method used in the DRS is given in, for example, \citet{gordon2012}. However, in the limit where read-noise is negligible and photon noise dominates, $\gamma \approx |\beta_{ij}|^2 + |\beta_{jk}|^2 + |\beta_{ik}|^2 - 2N$ where $\beta_{ij}$ is the photon bias used in Eq. \ref{eq:matisse_incoh_corrflux} for a baseline $\Vec{B}_{ij}$ and $N$ is the mean number of photons in the interferogram \citep{wirnitzer1985}. For bright sources, the $\gamma$ term is negligible. 

\begin{figure*}
    \centering
    \includegraphics[width=1.0\textwidth]{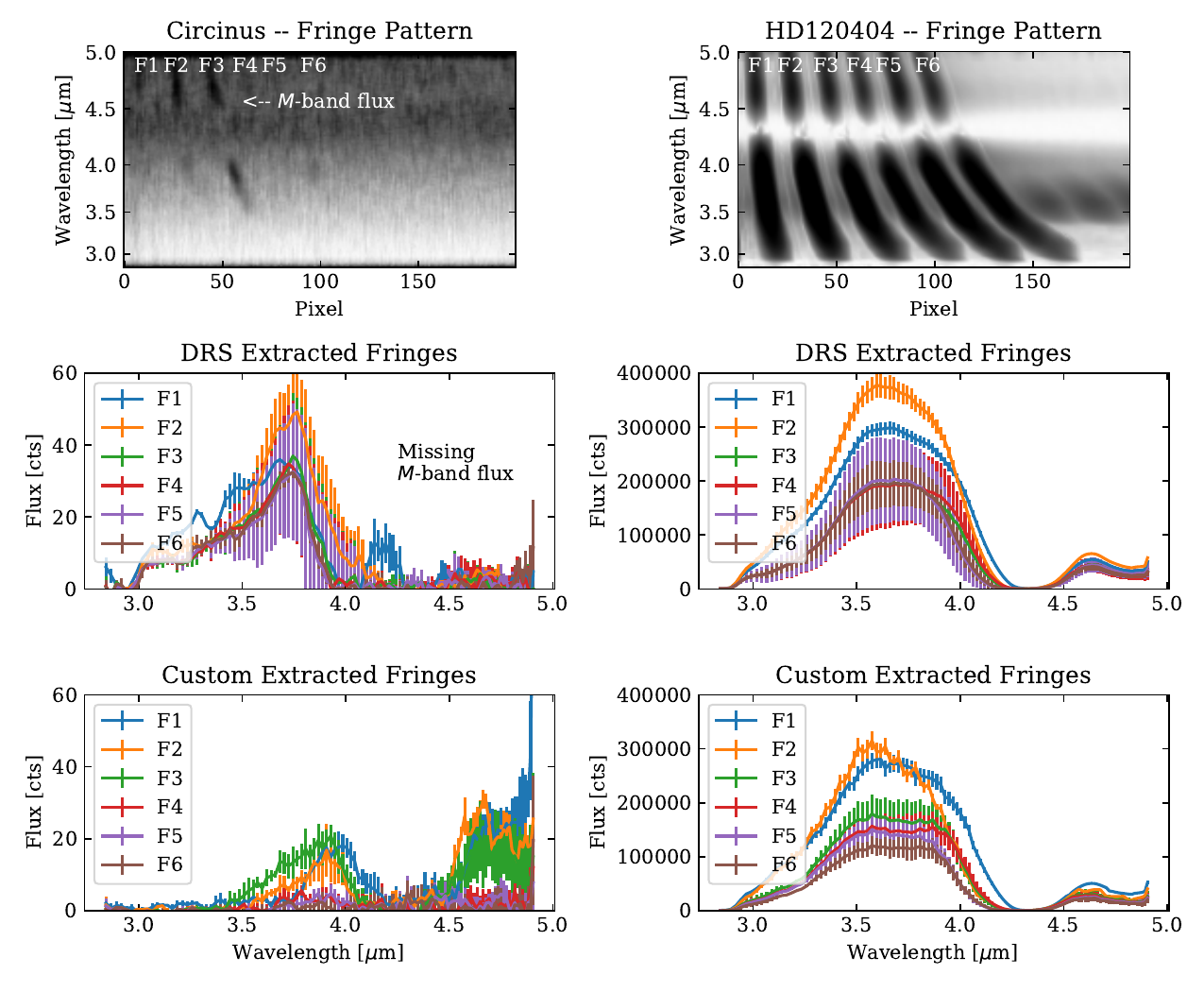}
    \caption[Comparison of DRS-extracted and custom-extracted correlated fluxes for Circinus (\textit{left}) and a calibration star, HD120404 (\textit{right}).]{Comparison of DRS-extracted and custom-extracted correlated fluxes for Circinus (\textit{left}) and a calibration star, HD120404 (\textit{right}). From top to bottom, the panels show the cleaned fringe pattern (the Fourier transform of the cleaned interferogram), the six fringes extracted with the DRS, and the same six fringes extracted with our method. In the Circinus fringe pattern, there is significant flux in the $M$ band that is missing in the DRS fringes but is present in our method.}
    \label{fig:circl_reduction}
\end{figure*}

\section{Reduced data}
\subsection{Correlated fluxes}
In Figs. \ref{fig:cflux0} and \ref{fig:cflux1} we present the $LM$-band correlated flux for each baseline, reduced and calibrated as described in Sect. \ref{sec:circ_obs}. 

\begin{figure*}
\renewcommand\thefigure{B.1}
    \centering
    \includegraphics[width=1.0\textwidth]{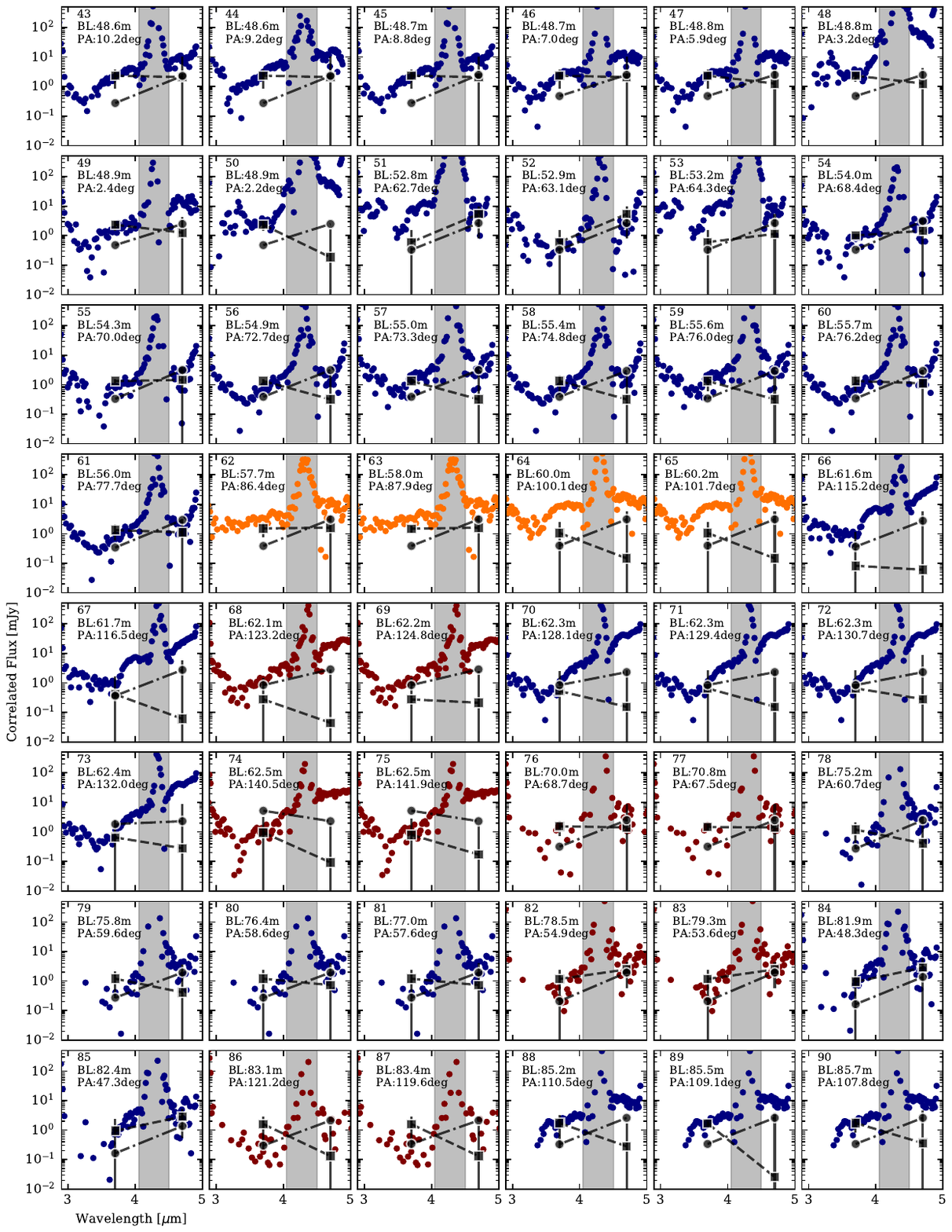}
    \caption[Continuation of Fig. \ref{fig:cflux0}]{Continuation of Fig. \ref{fig:cflux0}.}
    \label{fig:cflux1}
\end{figure*}
\begin{figure*}
\renewcommand\thefigure{B.1}
    \centering
    \ContinuedFloat
    \includegraphics[width=1.0\textwidth]{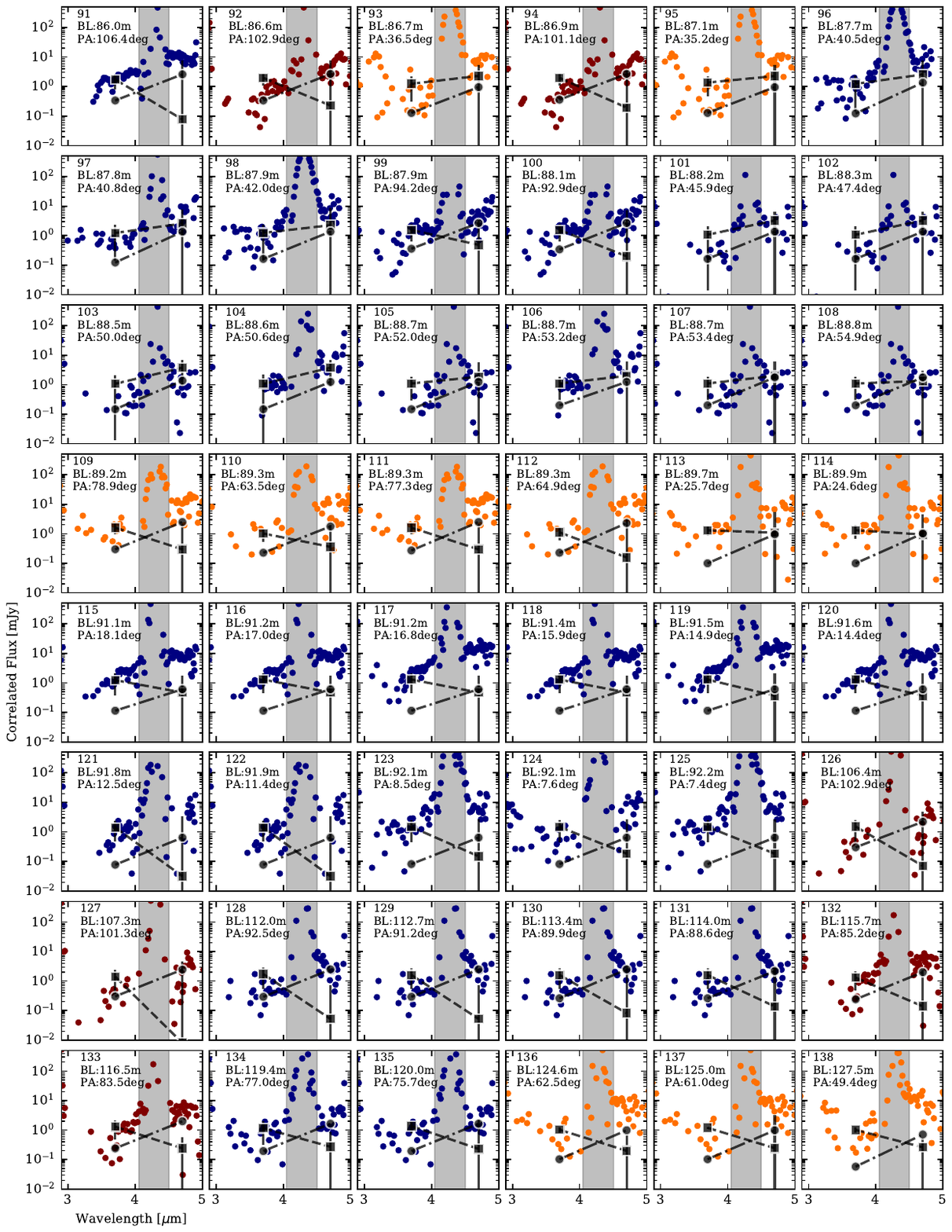}
    \caption{Continued.}
    \label{fig:cflux2}
\end{figure*}
\begin{figure*}
\renewcommand\thefigure{B.1}
    \centering
    \ContinuedFloat
    \includegraphics[width=1.0\textwidth]{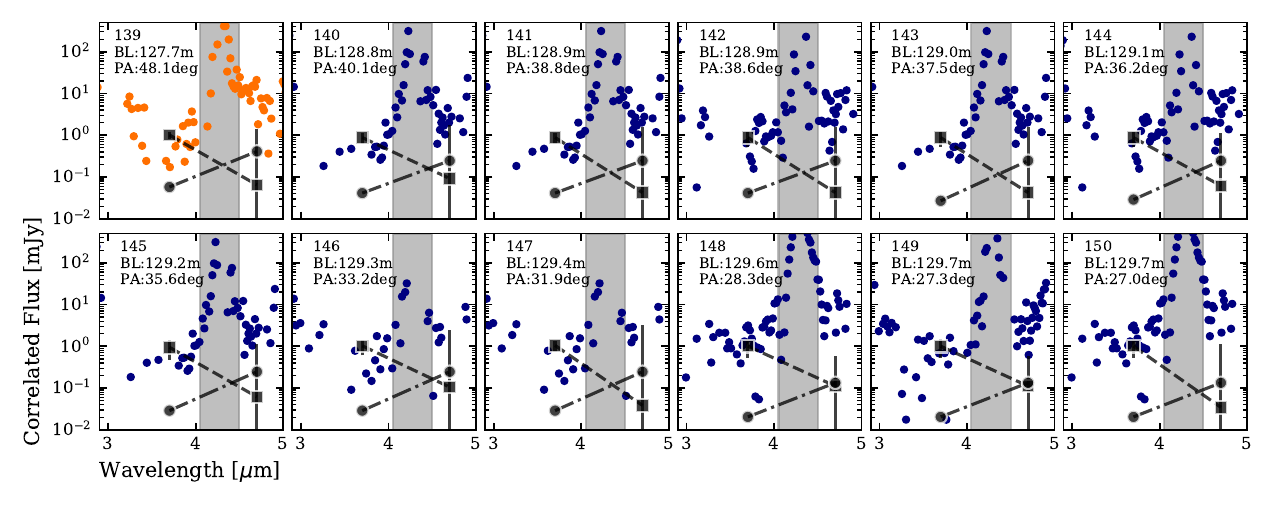}
    \caption{Continued.}
    \label{fig:cflux3}
\end{figure*}

\subsection{MATISSE $LM$-band closure phases}
In Fig. \ref{fig:t3phi0} we present the $LM$-band closure phase spectrum for each closure triangle, reduced and calibrated as described in Sect. \ref{sec:circ_obs}.

\begin{figure*}
\renewcommand\thefigure{B.2}
    \centering
    \includegraphics[width=0.98\textwidth]{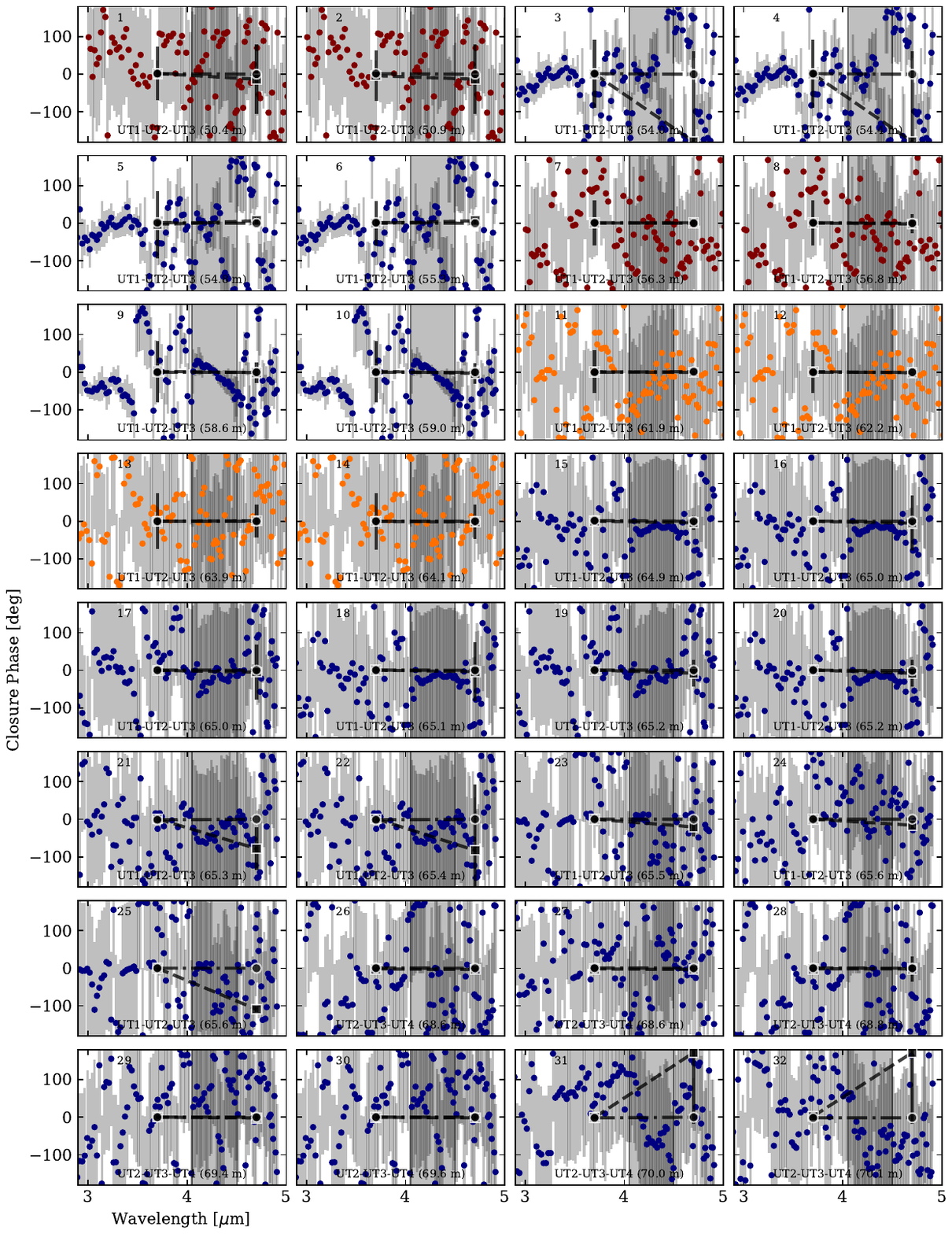}
    
    \caption[$LM$-band closure phase data for Circinus from March 2020 (blue), February 2021 (yellow), and May 2021 (red).]{\small{$LM$-band closure phase data for Circinus from March 2020 (blue), February 2021 (yellow), and May 2021 (red). Presented errors come from both the calibrator phase uncertainty and the statistical variation of the observables within a set of observing cycles. 
    }
    }
    \label{fig:t3phi0}
\end{figure*}
\begin{figure*}
\renewcommand\thefigure{B.2}
    \ContinuedFloat
    \centering
    \includegraphics[width=1.0\textwidth]{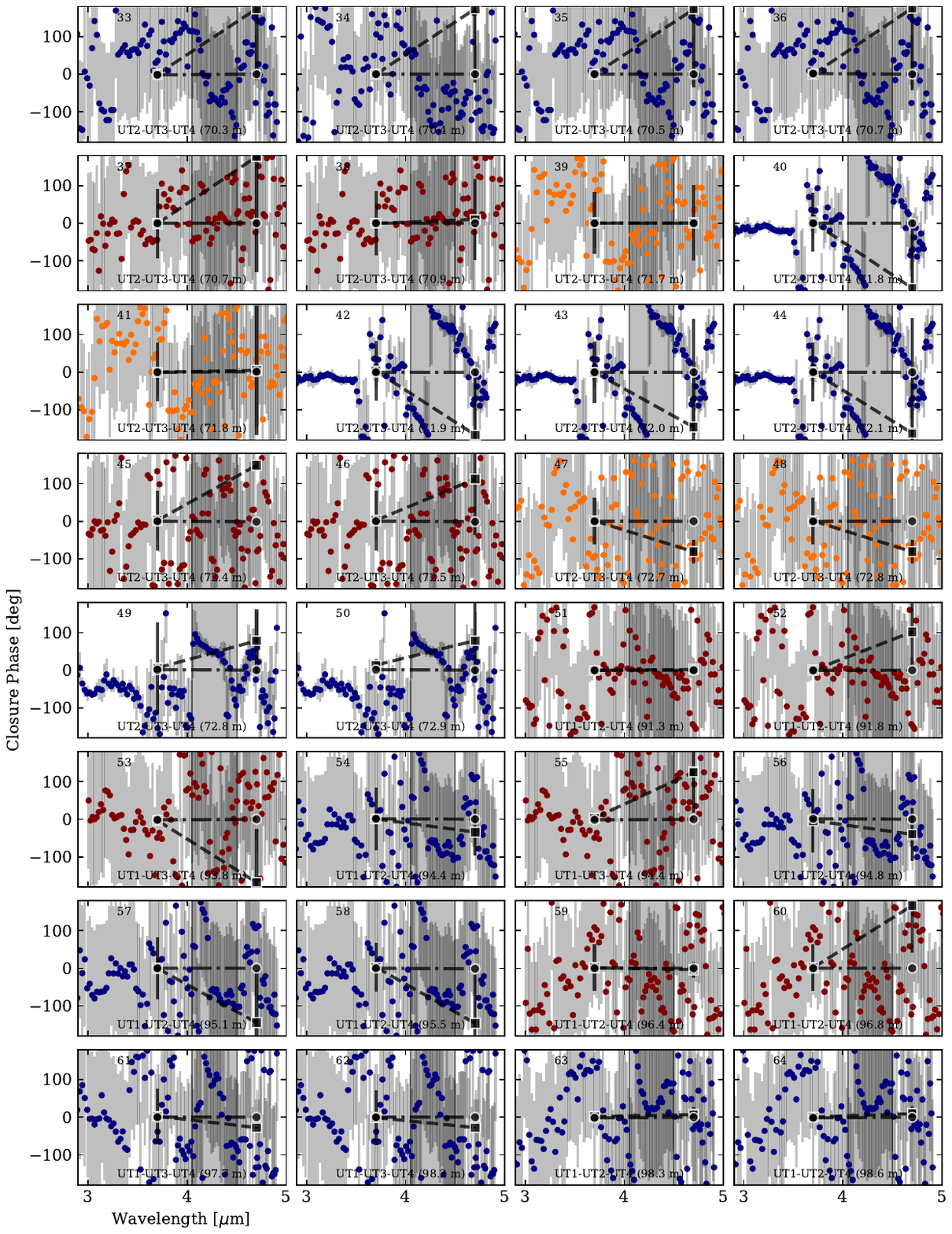}
    \caption{Continued.}
    \label{fig:t3phi1}
\end{figure*}
\begin{figure*}
\renewcommand\thefigure{B.2}
\ContinuedFloat
    \centering
    \includegraphics[width=1.0\textwidth]{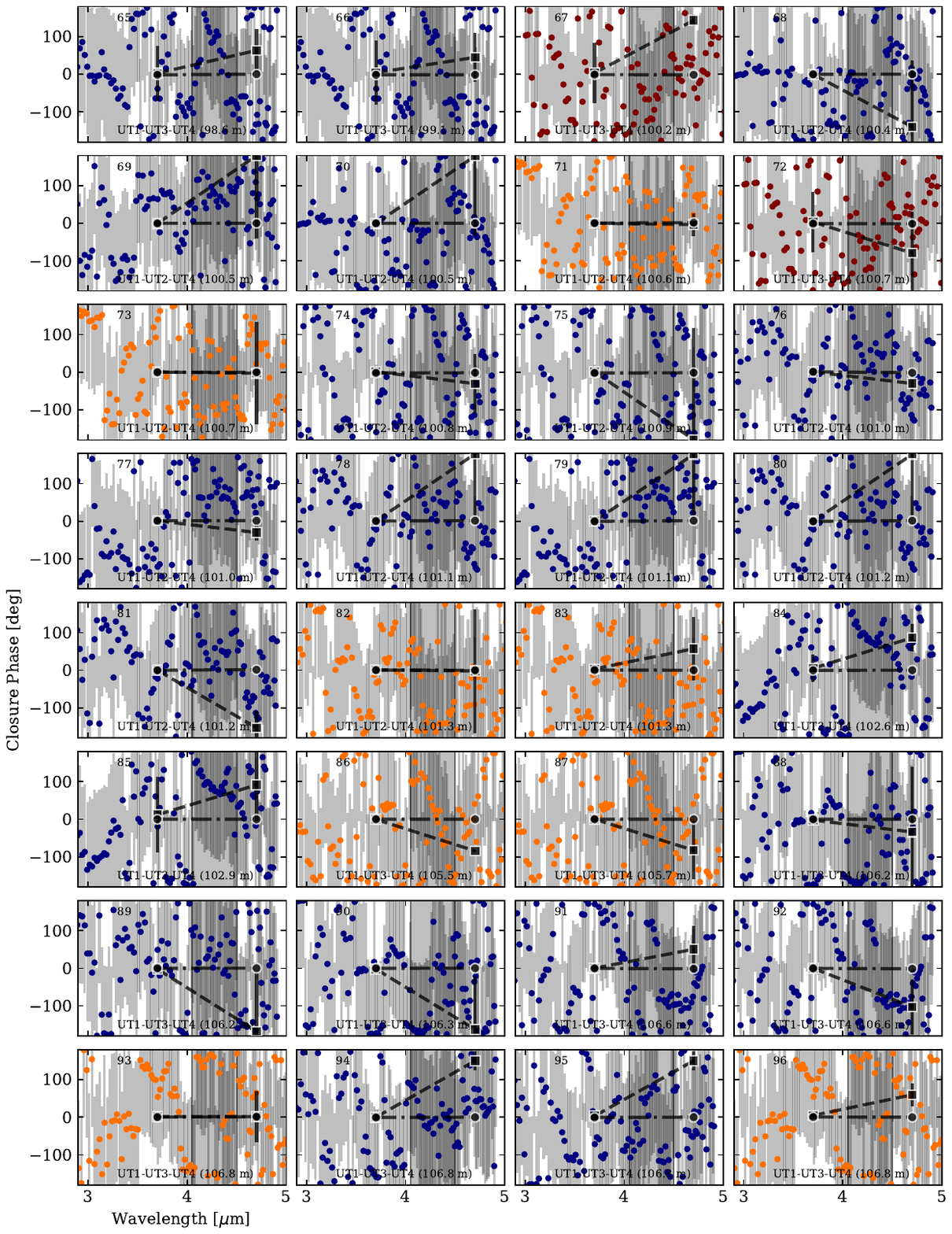}
    \caption{Continued.}
    \label{fig:t3phi2}
\end{figure*}
\begin{figure*}
\renewcommand\thefigure{B.2}
\ContinuedFloat
    \centering
    \includegraphics[width=1.0\textwidth]{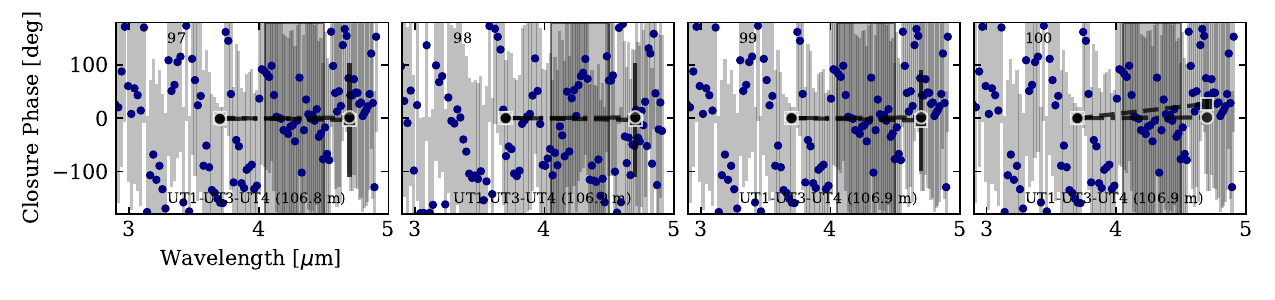}
    \caption{Continued.}
    \label{fig:t3phi3}
\end{figure*}

\section{Two-BB fits}
\begin{figure*}
\renewcommand\thefigure{C.1}
    \centering
    \includegraphics[width=0.44\textwidth]{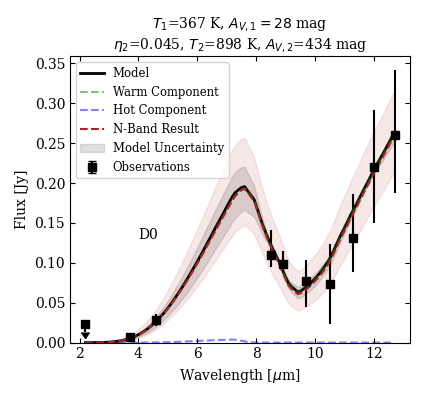}
    \includegraphics[width=0.44\textwidth]{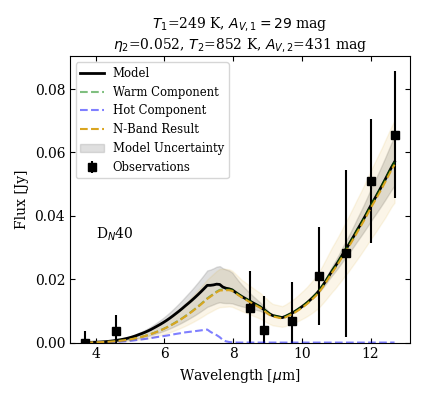}
    \includegraphics[width=0.44\textwidth]{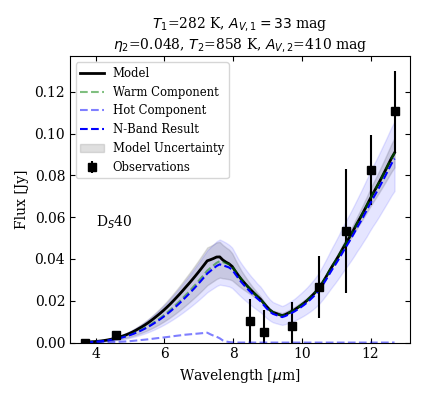} 
    
    \caption[One-blackbody (\textit{left}) and two-blackbody (\textit{right}) fits for the aperture-extracted Circinus \textit{LMN} fluxes.]{Two-BB fits for the aperture-extracted Circinus \textit{LMN} fluxes. The colors are the same as in Fig. \ref{fig:circl_immodel}, with D0 in red, D$_N$40 in yellow, and D$_S$40 in red. There is little discernible improvement in fit quality with the addition of a second component, and the second component is in all cases highly extincted. The fits using the $N$-band data alone are included for comparison. In aperture D0, the $K$-band measurement from \citet{burtscher2015} is included as an upper limit for the near-infrared flux. }
    \label{fig:circl_sedfit_2}
\end{figure*}

In Fig. \ref{fig:circl_sedfit_2} we show the resulting two-BB fits for the Gaussian models discussed in Sect. \ref{sec:bb}.

\begin{figure*}
    \centering
    \includegraphics[width=1.0\textwidth]{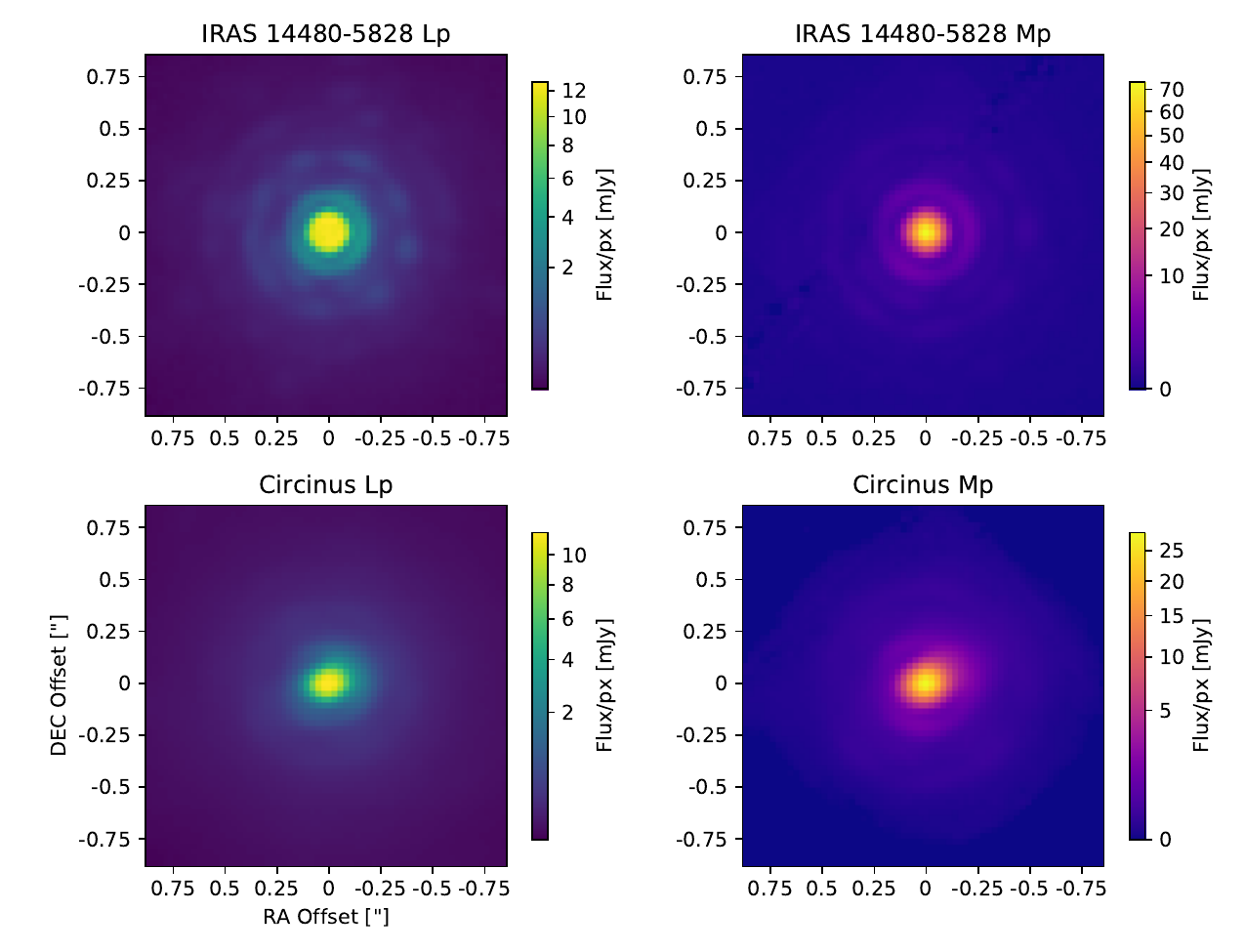}
    \caption{NACO flux-calibrated images for IRAS 14480-5828 and Circinus.}
    \label{fig:naco_im}
\end{figure*}

\begin{figure*}
    \centering
    \includegraphics[width=0.45\textwidth]{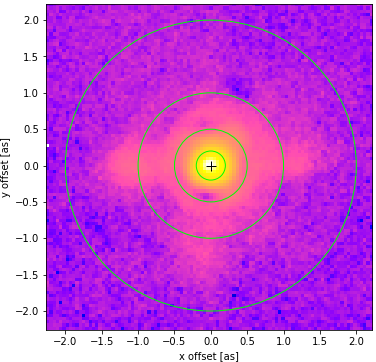}
    \caption{VISIR flux-calibrated M-band image of Circinus with 0.4", 1", 2", and 4" apertures overlaid. Here we clearly see the extended polar dust structures to the east and west of the nucleus.}
    \label{fig:visir_im}
\end{figure*}

\end{appendix}

\end{document}